\documentclass[11pt,a4paper]{article}
\usepackage[a4paper, total={6in, 8in}]{geometry}
\usepackage[utf8]{inputenc}
\usepackage{authblk}
\usepackage{hyperref}

\usepackage{ulem}
\usepackage{amsmath}
\usepackage{amsthm}
\usepackage{amssymb}
\usepackage{lmodern}
\usepackage{multirow}
\usepackage{graphicx}
\usepackage{subcaption}
\usepackage{placeins}
\usepackage{xcolor}

\usepackage{float}

\usepackage{cleveref}

\usepackage{cite}

\usepackage{lineno}
\usepackage{ulem}
\begin{document}

\normalem

\title{Finding new physics without learning about it: anomaly detection as a tool for searches at colliders}

\author[1]{M. Crispim Rom\~{a}o}
\author[1,2]{N. F. Castro}
\author[1]{R. Pedro}
\affil[1]{LIP, Av. Professor Gama Pinto 2, 1649-003 Lisboa, Portugal}
\affil[2]{Departamento de F\'{i}sica, Escola de Ci\^{e}ncias, Universidade do Minho, 4710-057 Braga, Portugal}

\maketitle

\begin{abstract} In this paper we propose a new strategy, based on anomaly detection methods, to search for new physics phenomena
    at colliders independently of the details of such new events. For this purpose, machine learning techniques are trained using
    Standard Model events, with the corresponding outputs being sensitive to physics beyond it. We explore three novel AD methods
    in HEP: Isolation Forest, Histogram Based Outlier Detection, and Deep Support Vector Data Description; alongside the most
    customary Autoencoder. In order to evaluate the sensitivity of the proposed approach, predictions from specific new physics
    models are considered and compared to those achieved when using fully supervised deep neural networks. A comparison between
    shallow and deep anomaly detection techniques is also presented. Our results demonstrate the potential of semi-supervised
    anomaly detection techniques to extensively explore the present and future hadron colliders' data.
\end{abstract}

\section{Introduction}

While the Standard Model of Particle Physics (SM) has been extremely successful in describing the experimental data accumulated so far, a significant number of open questions remains~\cite{Ellis:2012zz} and thus the search for new phenomena is a key aspect of the physics programme of present and future colliders. Given the practical difficulty of performing dedicated searches for all possible models and event topologies, inclusive searches and model-independent approaches are popular strategies to find a compromise between sensitivity and model independence of the experimental analyses. In fact, generic model-unspecific searches were conducted in the past by the D0~\cite{Abbott:2000gx,Abbott_2001}, CDF~\cite{Aaltonen_2008,Aaltonen_2009} and H1~\cite{Aktas:2004pz,Aaron:2008aa} experiments at the Tevatron and HERA, respectively, and are also performed nowadays by the ATLAS~\cite{Aaboud_2019} and CMS~\cite{cmscollaboration2020music} Collaborations of the Large Hadron Collider. Nonetheless, there is always the concern that a possible signal beyond the SM (BSM) is missed simply because the adopted strategy is not sensitive to it. In a previous work~\cite{TransfDL} we demonstrated that a possible direction to improve the sensitivity to BSM events without depending too much on the details of the considered signals is the supervised training of deep neural networks (DNN) since the performance of these networks does not significantly degrade when they are applied to another signal than the one used for training, as long as these signals are not very different from a topological point of view. A step forward in this direction is the use of anomaly detection (AD) methods, where only SM events are used in the training of the machine learning algorithm, allowing to isolate any BSM signal without knowing their details, avoiding any prior dependence and bias on the new physics that we are trying to discover.

The AD approach relies on identifying abnormal events in a data sample consisting, in the majority or completely, of normal events belonging to the same class. The problem is usually addressed by unsupervised learning with classical shallow algorithms running to identify the outlier events. In deep learning, Artificial Neural Networks such as autoencoders (AE) have found their use as anomaly detectors since the error on the reconstruction of the inputs given by a model trained exclusively on normal events can be interpreted as an anomaly score. A known drawback of typical shallow methods, such as One-Class Support Vector Machines (OC-SVM), is the failure for high-dimensional data with many entries. This leads to a need for substantial feature engineering and dimensionality reduction before their application. On the other hand, the deep learning architecture of the AE family deals well with high-dimensional data and performs in anomaly detection despite not being trained specifically for discerning outlier events in the data.

The potential to isolate any unexpected signal from the SM prediction, commonly referred to as background, has motivated a growing interest for AD in HEP. For example, in ref.~\cite{Collins_2018}, an unsupervised bump hunting approach using CWoLa~\cite{Metodiev_2017} is proposed, while in ref.~\cite{De_Simone_2019}, a Machine Learning (ML) model based on k-Nearest Neighbours is used to estimate event densities and assess how likely a new event is. Ref.~\cite{D_Agnolo_2019} employs Neural Networks to compare the distribution of two samples and derive statistical tests to evaluate if any new physics is present. In refs.~\cite{Cerri_2019, Farina_2020, Blance_2019} three different AE produce distributions of reconstruction errors to be used as anomaly scores, whereas~\cite{Hajer_2020} conjugates an AE with a Linear Outlier Factor. More recently, in~\cite{Nachman_2020, Andreassen_2020}, novel non-ML approaches using density estimates are employed. {On top of these examples, we also refer to the growing literature on the application of unsupervised or weakly supervised methods used to further understand the data generated at colliders~\cite{aguilar2017generic, heimel2019qcd, dillon2019uncovering, d2019learning, collins2019extending, amram2020tag, dillon2020learning, atlas2020dijet,knapp2020adversarially}.}

The search for outlier events using anomaly detection techniques has a vast potential in the search for new phenomena in colliders, both at trigger (\emph{i.e.} online) and analysis (\emph{i.e.} offline) levels. Both applications have particular challenges and require dedicated efforts, namely in terms of the background modeling, event rates and statistical interpretation of the results. In this paper, we present three new unsupervised ML models for AD in the context of the offline analysis of HEP collisions, in addition to an AE, contributing to the path towards the use of such techniques by the experimental collaborations. In order to test their sensitivity to different BSM signals, the signals considered in~\cite{TransfDL} are used as benchmarks to access the performance of the proposed approach by comparing it with supervised DNN classifiers trained on the same signals. In this way, we compare the performance of the AD methods to supervised DNNs. As such, {we further contribute to the ongoing effort -- see for example~\cite{Cerri_2019, knapp2020adversarially} -- to systematically compare} different unsupervised AD methods in searches for new physics.

\section{Methods for anomaly detection}

We use shallow and deep learning techniques trained on a data sample of Standard Model simulated events and test the ability of each model to identify new physics events with benchmark signals unseen during the training phase. Histogram-based outlier detection (HBOS)~\cite{Goldstein2012HistogrambasedOS} and Isolation Forest (iForest)~\cite{10.1109/ICDM.2008.17} are the shallow models explored. These methods are guided to isolate instances of the data in the tails of the feature distributions and, unlike OC-SVM, are fast and scalable to high-dimensional data with many instances. As a deep model, we analyse the recently proposed Deep Support Vector Data Description (Deep SVDD)~\cite{pmlr-v80-ruff18a}. Contrary to an AE, the Deep SVDD is designed for outlier discovery. AEs, popularly used in AD tasks, are also explored.

\subsection{Histogram-based outlier detection}

In HBOS, a histogram is computed for each input feature and an anomaly score 
is derived based on how populated the bins where an instance falls on are. 
In the training phase, the predicted SM yields are used to construct the 
bins. On the test phase, the score of a new instance is computed as follows. 
For each of its features, we see in what bin of the histogram its value 
falls on, and assign an associated score of $\log_2(\text{Hist})$, with Hist 
being the density of the histogram where the instance value of the feature is, \emph{i.e.} the height of the bin that contains that value. The total 
anomaly score is the sum across all features.

\subsection{Isolation forest}

The iForest algorithm randomly selects an input feature and a split value within the feature boundaries to recursively partition the data. The idea is that outliers are easier to isolate than normal instances of the data and the number of data splits can be used as a base for an anomaly score. In the training phase, the iForest model learns the feature boundaries from the training sample and on the test phase each event is isolated and its outlyingness is obtained.

\subsection{Deep autoencoder}

Deep AE is a deep architecture that learns to compress (encode) and then decompress (decode) data through a bottleneck intermediate layer that has a smaller dimensionality than the data. The AE is trained by minimising the reconstruction error, \emph{i.e.} how different a decoded instance is from the original, through the training objective:
\begin{equation}\label{eq:aeloss}
    \min_{\mathcal{W}} \frac{1}{n}\sum_i || \text{AE}(\mathbf{x}_i, \mathcal{W}) - \mathbf{x}_i ||^2 \ ,
\end{equation}
where $\mathcal{W}$ are the weights of the AE, $\mathbf{x}_i$ the feature vector of the $i$th event and $n$ the total number of events. Since uncommon events will, in principle, be harder to reconstruct than more common ones, the reconstruction error can then be used as an anomaly score.

\subsection{Deep support vector data description}

The Deep SVDD architecture is designed in analogy to its shallow counterpart, the Support Vector Data Description, which in turn is closely related to OC-SVM. In SVDD, the data is mapped into an abstract feature space and, during training, we minimise the mean distance of data points to the centre of the data distribution in this space. In the deep version, this is implemented as follows. We initialise a DNN and calculate the average position of its outputs given the training set. This will give us the centre of the distribution of the data in the space defined by the last layer of the DNN. Training is then performed as to minimise the distance of all points of the training set to this centre and can be expressed through the training objective:
\begin{equation}\label{eq:dsvddloss}
    \min_{\mathcal{W}} \frac{1}{n} \sum_i || \text{DNN}(\mathbf{x}_i, \mathcal{W}) - \mathbf{c} || ^2 \ ,
\end{equation}
where $\mathcal{W}$ are the weights of the DNN, $\mathbf{c}$ the centre of the distribution in the output space, $\mathbf{x}_i$ the feature vector of the $i$th event. In order to prevent pathological behaviours arising from trivial solutions associated with collapses of the whole distribution to $\mathbf{c}$, the DNN must have non-saturated activation functions, it must not have bias terms, and $\mathbf{c}$ can be neither the origin of the output space nor a learnable parameter. The anomaly score of an event in a Deep SVDD is then deduced from how far from the centre, $\mathbf{c}$, the event lies.

\subsection{Supervised classifier}

In addition, we trained a supervised classifier, based on deep neural networks, for each benchmark signal (\emph{c.f.} \cref{samples}). This will provide us with a baseline with which to compare the AD algorithms performance.

\section{Simulated datasets}
\label{samples}

We tested the different AD methods in the context of collider searches and our dataset is composed of simulated proton-proton collision events~\cite{zenodo5126747}. The samples were generated with MADGRAPH5\_MCATNLO 2.6.5~\cite{madgraph} at leading order with a collision centre-of-mass energy of 13~TeV. Pythia 8.2~\cite{pythia} was employed to simulate the parton shower and hadronisation, with the CMS CUETP8M1~\cite{CUETP8M1} underlying event tuning and the NNPDF 2.3~\cite{NNPDF2.3} parton distribution functions. The detection of the collision products was accomplished with a multipurpose detector simulator, Delphes 3~\cite{delphes}. The configuration of Delphes was kept to the default, matching the parameters of the CMS detector. Jets and large-radius jets are reconstructed using the anti-$\kappa_t$ algorithm~\cite{Cacciari:2008gp} with a radius parameter of $R =0.5$ and 0.8, respectively.

One of our goals is to compare the AD performance to the one obtained with dedicated supervised deep learning, which we explored previously~\cite{TransfDL}. For this reason, we studied the same BSM signals, namely the pair production of vector-like $T$-quarks (either produced via SM gluons~\cite{TopPartners} or BSM heavy gluons~\cite{HGinterpretation}) and $tZ$ production through a flavour changing neutral current (FCNC) vertex~\cite{FCNCmodel2}. In total, seven benchmark signals were generated: $T\bar T$ with $m_T={1.0,1.2,1.4}$~TeV produced via SM gluon or a massive 3~TeV gluon, and $tZ$ FCNC production.

We preselected events broadly compatible with the signal topologies commonly considered by the ATLAS and CMS experiments~\cite{ATLASZtag, CMSZtag, ATLASFCNC, CMSFCNC}: at least two final state leptons (\emph{i.e.} electrons or muons), at least one $b$-tagged jet, and large scalar sum of transverse momentum ($p_T$) of all reconstructed particles in the event ($H_T>500$~GeV)\footnote{The transverse plane is defined with respect to the proton colliding beams.}. The most important SM processes compatible with the event selection topology are $Z$+jets, top pair ($t\bar{t}$) production and dibosons ($WW$, $WZ$ and $ZZ$). The generation of each of these processes was sampled in kinematic regions to ensure a good statistical representation across the entire phase space, and especially in the tails of the distributions, where anomalous events are particularly expected. This sampling employed event generation filters at parton level according to:
\begin{itemize}
    \item The top/anti-top $p_T$ ($p_T^{top}$) for $t\bar{t}$: $p_T^{top}<100$~GeV, $p_T^{top}\in[100,250]$~GeV, $p_T^{top}>250$~GeV;
    \item The scalar sum of the $p_T$ of the hard-scatter outgoing particles for $Z$+jets: $S_T<250$~GeV, $S_T\in[250,500]$~GeV, $S_T>500$~GeV;
    \item $W/Z$ $p_T$ ($p_T^{W/Z}$) for dibosons: $p_T^{W/Z}<250$~GeV, $p_T^{W/Z}\in[250,500]$~GeV, $p_T^{W/Z}>500$~GeV.
\end{itemize}

In order to ensure a reasonable statistics across the relevant phase space, the $Z$+jets simulation was further split into the jet flavour as $Zjj$ and $Zbb$. Over 18~M events were simulated: 500~k per signal sample, 8~M for $Z+$jets, 3~M for $t\bar{t}$ and 1.5~M per diboson sample.

Furthermore, the generated events were also hadronised with Herwig~7~\cite{herwig++, herwig7}, employing NNPDF 2.3~\cite{NNPDF2.3} parton distribution functions, in order to produce an alternative set of samples to test the robustness of AD techniques against uncertainties on the parton shower and hadronisation modelling.

The SM cocktail used to train the AD methods is composed of the SM simulated samples, each normalised to the expected yield after selection using the generation cross-section at leading order, computed with MADGRAPH5, and matched to a target luminosity of 150~fb$^{-1}$. This normalisation is parsed as a form of event weights to the AD method. The data features correspond to basic information constituted of the four-momenta of the reconstructed particles as provided by the Delphes simulation:
\begin{itemize}
    \item $(\eta,\phi,p_T,m)$ of the 5 leading jets and large-radius jets;
    \item $(\eta,\phi,p_T)$ of the 2 leading electrons and muons;
    \item multiplicity of jets, large-radius jets, electrons and muons;
    \item $(E_T,\phi)$ of the missing transverse energy.
\end{itemize}

Some of these features manifest an accumulation of density at the origin. This happens for objects that might not have been reconstructed, such as sub-leading large-radius jets or flavour-explicit leptons. This will produce density functions for these features, which are not continuous and can hinder the performance of deep learning models. In light of Universal Approximation Theorems for neural networks~\cite{hornik1989multilayer,cybenko1989approximation,hornik1991approximation,lu2017expressive}, we know that neural networks are only guaranteed to approximate any \emph{continuous} function when given enough capacity, \emph{i.e.} enough width and/or units. Therefore, it is only reasonable to assume that when the features are described by non-continuous densities, a neural network will have to learn a non-continuous function during training that will be difficult to learn as it is not guaranteed that it can be approximated. Consequently, we prepared the data with a second set of features that aims to mitigate this. This second set of features, which we refer to as \emph{sanitised}, retains only the events with one large-radius jet while dropping the features of all sub-leading large-radius jets. In addition, we keep only the two leading leptons regardless of the flavour, dropping the remainder.

\section{Implementation details and training}

The data were split into train, validation and test sets with equal proportions to guarantee similar statistical representativity at each stage. When hyperparameters were tuned, the metrics used to help choosing the best configuration were computed on the validation set. A statistically independent test set was used to evaluate the performance of the AD methods in isolating BSM signals.

\subsection{Shallow methods}

%\paragraph{HBOS}

We implemented the HBOS algorithm based on the {\texttt{pyod}} Python toolkit~\cite{zhao2019pyod}, but we changed the code to take sample normalisation weights into account when computing the histograms. For the iForest, we based our implementation on the Scikit-Learn~\cite{scikit-learn} through the {\texttt{pyod}} wrapper~\cite{zhao2019pyod}.

For both the HBOS and the iForest implementations the data was preprocessed by a standardisation step, which sets all the features means to 0 and their standard deviation to unity, followed by a principal component rotation, where we retained the full dimensionality of the feature space. The purpose of this rotation is to remove linear correlations between the features, an assumption that is required by these methods. The preprocessing steps were implemented with Scikit-Learn~\cite{scikit-learn}.

\subsection{Deep methods}

%\paragraph{AE} 

We implemented the deep models in TensorFlow 2.3~\cite{TF}. {In order to find the best hyperparameters for the deep architectures, we implemented a bayesian hyperparameter optimisation step using the Python package {\texttt{optuna}~\cite{akiba2019optuna}}. The hyperparameter optimisation step made use of the \texttt{optuna} built-in Tree-Structured Parzen Estimator~\cite{bergstra2011algorithms} to suggest new hyperparameter combinations, over a loop of increasing number of maximum epochs to improve search efficiency. In addition, manually discovered promising hyperparameter combinations were added to the evaluation queue.}

{A crucial hyperparameter to be fixed before the hyperparameter optimisation loop is that of the dimensionality of the latent space of the AE and the embedding space of the Deep SVDD. The reason to fix it is twofold: on the one hand an AE hyperparameter optimisation step will always prefer a large latent space, which will fix it to the highest value possible during search; on the other hand, it is difficult to compare distributions of distances on different dimensions, making model comparison and selection for the Deep SVDD challenging. In~\cite{pmlr-v80-ruff18a} the second problem was circumvented by reusing the AE encoder as a pre-trained Deep SVDD. In this work, we let the Deep SVDD to be trained from scratch, but fixed the embedding dimension to be the same as of the latent space of the AE. We did not use the encoder of a trained AE as we observed that this led to instabilities during training and difficulties in reproducing the same results. Instead, we fixed the embedding dimension and optimised the remainder hyperparameters of the Deep SVDD using the same Bayesian search. The latent space dimension of the AE and the embedding space dimension of the Deep SVDD was set to $16$, as it is roughly a quarter of the input dimensionality.}

{For the Deep SVDD, the vector $\mathbf{c}$, which represents the centre of the distribution of the data in the embedding space, was calculated as follows. First, we defined the model and initialise all its learning parameteres. Next, and before any optimiser step, we forward pass the whole training set through this network and calculate the weighted average of the outputs. This will then be the centre of mass of the distribution in the embedding space and therefore sets $\mathbf{c}$.}

{All deep models were trained with a custom cosine-cyclical learning rate with warmup. The warmup phase was set to a $25$ epochs period, where the learning rate linearly increased from an initial value, \texttt{Initial LR}, to its maximum value, \texttt{Max LR}, both to be optimised during the optuna loop. The cycle was set with a period of $50$ epochs, during each period the learning rate oscillates between the maximum learning rate down to an order of magnitude lower. During the cycle phase, the maximum learning rate was multiplied by a factor, \texttt{gamma}, at the end of each epoch, exponentially decreasing it, which was optimised during the bayesian optimisation loop. We found this type of learning rate to significantly improve the converge speed of both AE and Deep SVDD, as well as to improve the training stability in terms of reproducibility of the final outcome. The training was stopped if no improvement of the loss on the validation set was observed for 200 epochs for the AE, 300 for the Deep SVDD, and 100 for the supervised classifiers, after which the weights of the best epoch were kept, persisting the best models at every stage. \footnote{We allowed a larger patience for the Deep SVDD early stop criteria as we observed the loss to oscillate significantly at early stages.} The AE was trained using mini-batches of size 4096, while the Deep SVDD and the supervised classifiers were trained in mini-batches of size 1024. All hidden layers activation functions were set to \texttt{LeakyReLu}.}

{In addition, all models were trained with the Adam optimiser~\cite{kingma2014adam}, through the weight-decay wrapper provided by Tensorflow-Addons in order to implement weight-decay regularisation compatible with Adam\cite{loshchilov2017decoupled}. The value of the weight-decay was optimised during the hyperparameter search loop. Furthermore, since the the Deep-SVDD cannot have non-homogeneous learnable parameters, i.e. biases, we implemented a non-trainable Batch Normalisation or otherwise the learnable mean would effectively behave as a bias term and lead to trivial collapse solutions. Since preventing trivial solutions requires not using saturating activation functions and learnable batch normalisation layers, one would expect only shallower networks to be successfully trained in order to avoid vanishing and exploding gradients. To mitigate this, we allowed for the gradients to be norm-clipped to a value to be optimised.}

{The hyperparameter optimisation loop details can be found in table~\cref{tab:hpsearch}. The best combinations were chosen by minimising validation loss, and the final configurations for the AD models for both feature sets can be seen in table~\cref{tab:hpbest}. We do not present the best hyperparameters for the supervised classifiers for brevity.}

\begin{table}[]
    \caption{Hyperparameter search spaces. The sampling for \texttt{Initial LR} and \texttt{Max LR} was performed logarithmically. For the AE the number of layers corresponds to both the number of encoder and decoder layers.}
    \label{tab:hpsearch}
    \begin{center}
        \begin{tabular}{l|l}
            Hyperparameter   & Possible Values                                                        \\ \hline
            Number of Layers & $[1,5]$                                                                \\
            Number of Units  & $[32, 256]$                                                            \\
            Initial LR       & $[10^{-8}, 10^{-3}]$                                                   \\
            Max LR           & $[10^{-3},10^{-1}]$                                                    \\
            Gamma            & $[0.95, 0.999]$ in steps of $0.001$                                    \\
            Weight Decay     & $\{0, 10^{-9}, 10^{-8}, 10^{-7}, 10^{-6}, 10^{-5}, 10^{-4}, 10^{-3}\}$ \\
            Clipnorm         & $\{\texttt{None},0.001, 0.01, 0.1, 1.0, 10.0, 100.0\}$
        \end{tabular}
    \end{center}
\end{table}

\begin{table}[]
    \caption{Best hyperparameter configurations for both deep AD models on both feature sets.}
    \label{tab:hpbest}
    \begin{center}
        \begin{tabular}{l|ll|ll}
                             & \multicolumn{2}{c|}{AE}           & \multicolumn{2}{c}{Deep SVDD}                                                                                        \\
            Hyperparameter   & \multicolumn{1}{c}{Full Features} & \multicolumn{1}{c|}{Sanitised Features} & \multicolumn{1}{c}{Full Features} & \multicolumn{1}{c}{Sanitised Features} \\ \hline
            Number of Layers & $5$                               & $3$                                     & $2$                               & $1$                                    \\
            Number of Units  & $171$                             & $93$                                    & $47$                              & $128$                                  \\
            Initial LR       & $7\times 10^{-6}$                 & $4.487459\times 10^{-7}$                & $5.834093\times10^{-8}$           & $10^{-6}$                              \\
            Max LR           & $0.023328$                        & $0.063960$                              & $0.005186$                        & $0.02$                                 \\
            Gamma            & $0.998$                           & $0.992$                                 & $0.971$                           & $0.995$                                \\
            Weight Decay     & $10^{-6}$                         & $0.0$                                   & $10^{-9}$                         & $10^{-8}$                              \\
            Clipnorm         & $0.10$                            & $100.0$                                 & $100.0$                           & $\texttt{None}$
        \end{tabular}
    \end{center}
\end{table}

%\paragraph{Deep SVDD}

For both the AE and the Deep SVDD methods, the anomaly score was derived from the loss, \emph{i.e.}~\cref{eq:aeloss,eq:dsvddloss}, by taking the base 10 logarithm of the values and scaling them as to fall in the interval $[0,1]$. {It is also important to reiterate that the signals samples were not used at any stage of both hyperparameters selection and AD model training.}

%\paragraph{Supervised Classifiers} 

% For each of the considered benchmark signals, we also trained a DNN to perform a binary classification between signal and SM background prediction. The hyperparameters used were the same for all cases and are shown in~\cref{tab:clfhp}. Each classifier was trained with the same learning rate cycle as the one used with the AE and Deep SVDD, on mini-batches of size 4096. The training was interrupted if no improvement in the validation set was observed for 200 epochs.

\subsection{Feature impact on reconstructions}

\begin{figure}[]
    \begin{subfigure}[]{\textwidth}\includegraphics[width=0.49\textwidth]{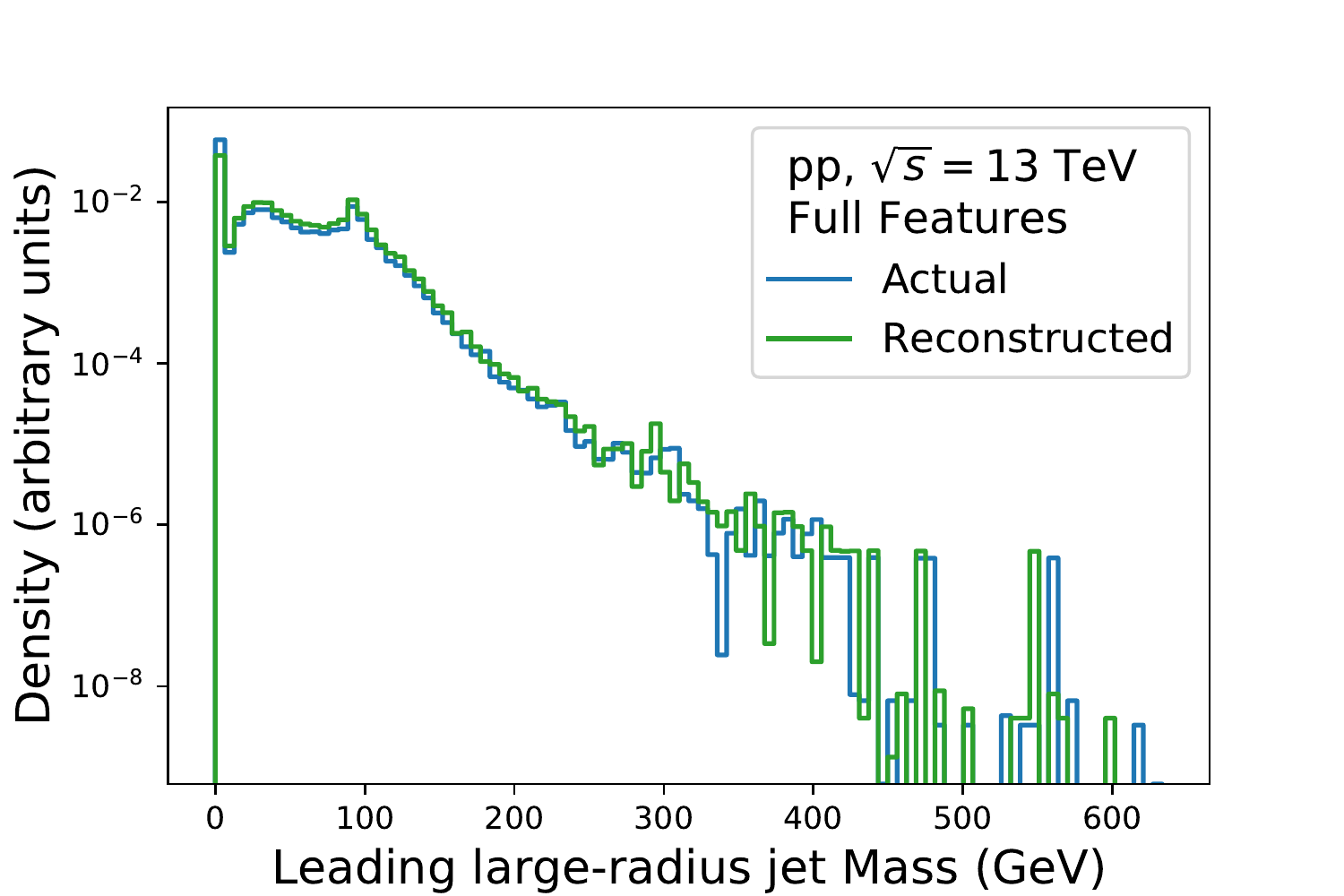} \includegraphics[width=0.49\textwidth]{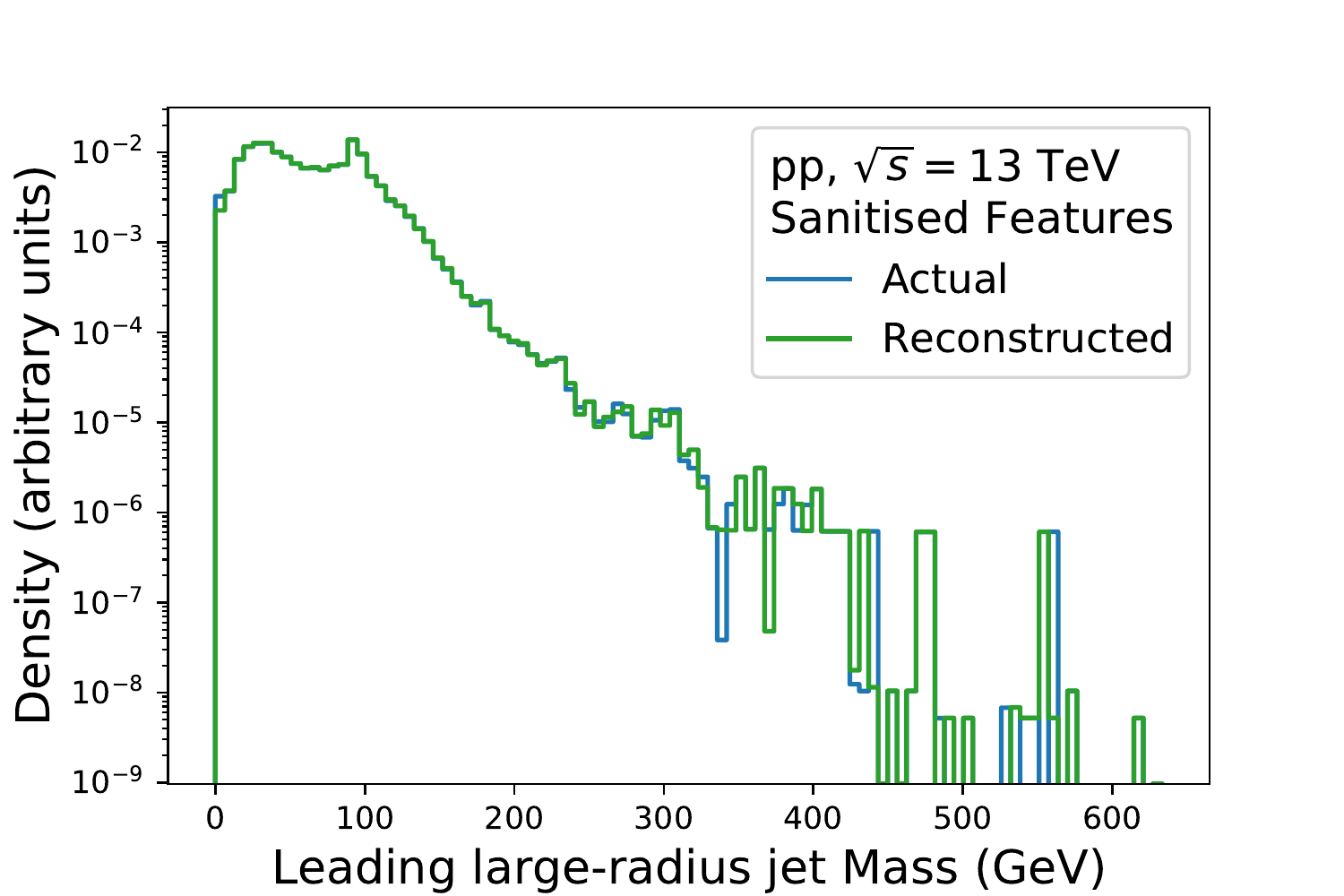}\end{subfigure}
    \begin{subfigure}[]{\textwidth}\includegraphics[width=0.49\textwidth]{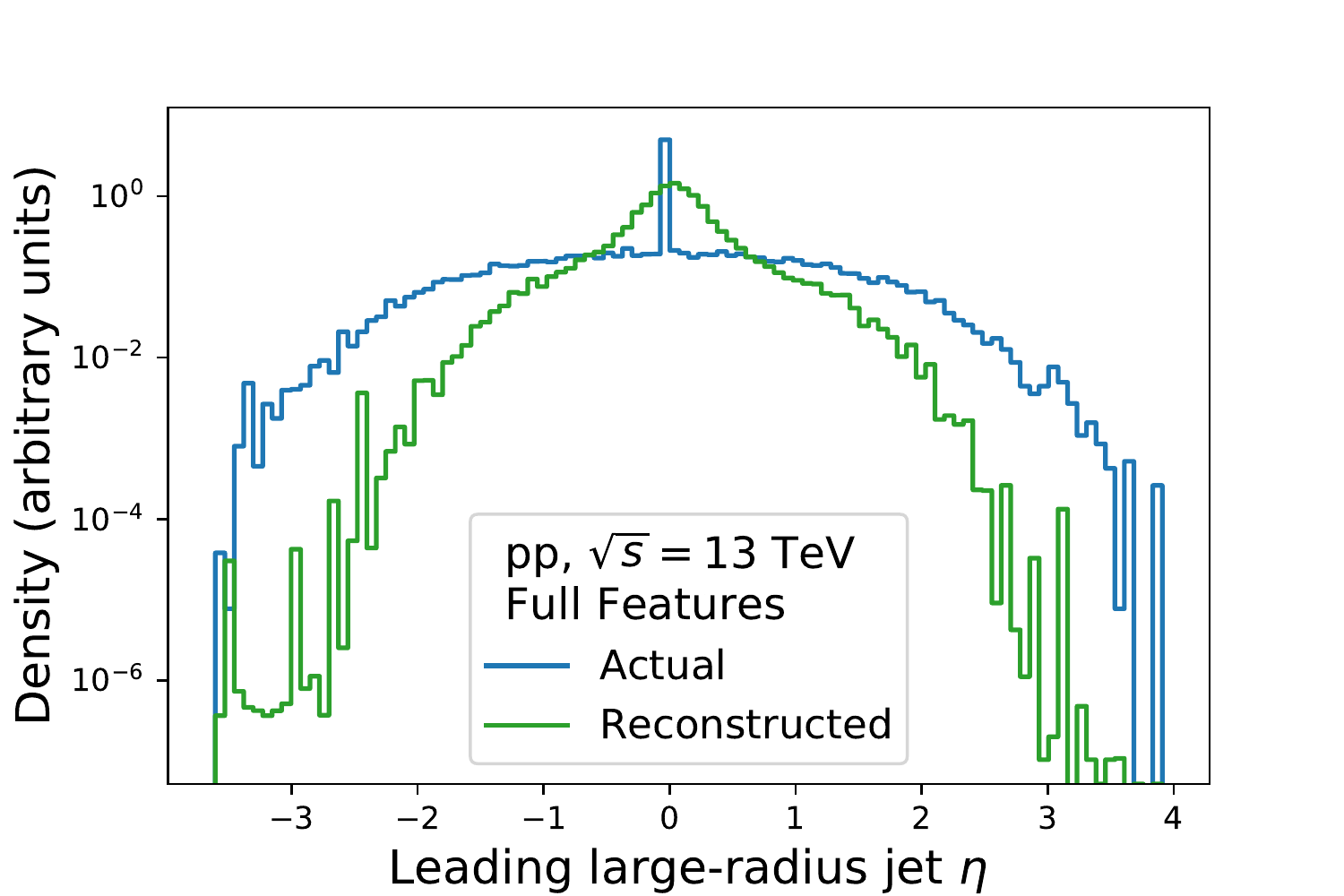} \includegraphics[width=0.49\textwidth]{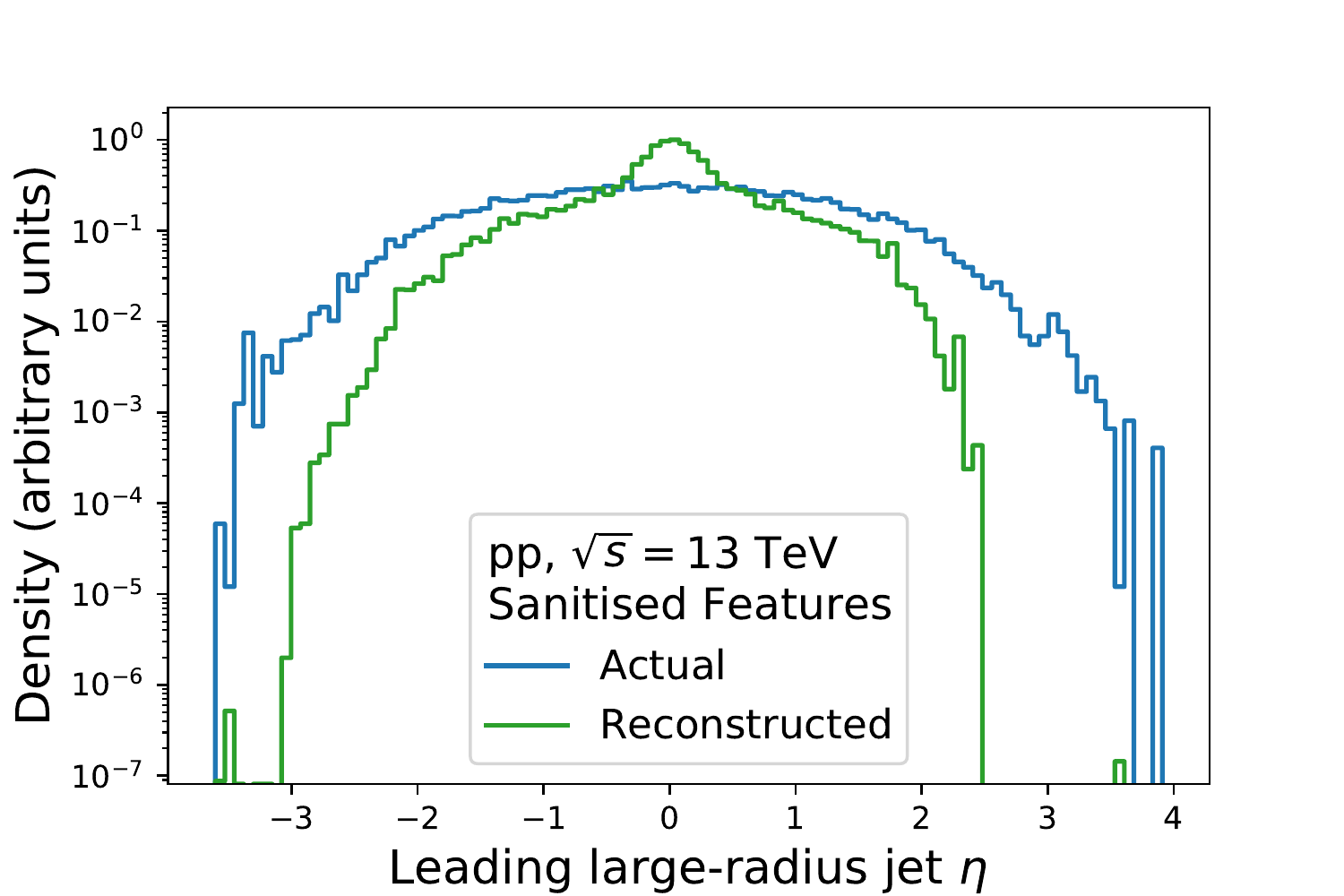}\end{subfigure}
    \begin{subfigure}[]{0.49\textwidth}\includegraphics[width=\textwidth]{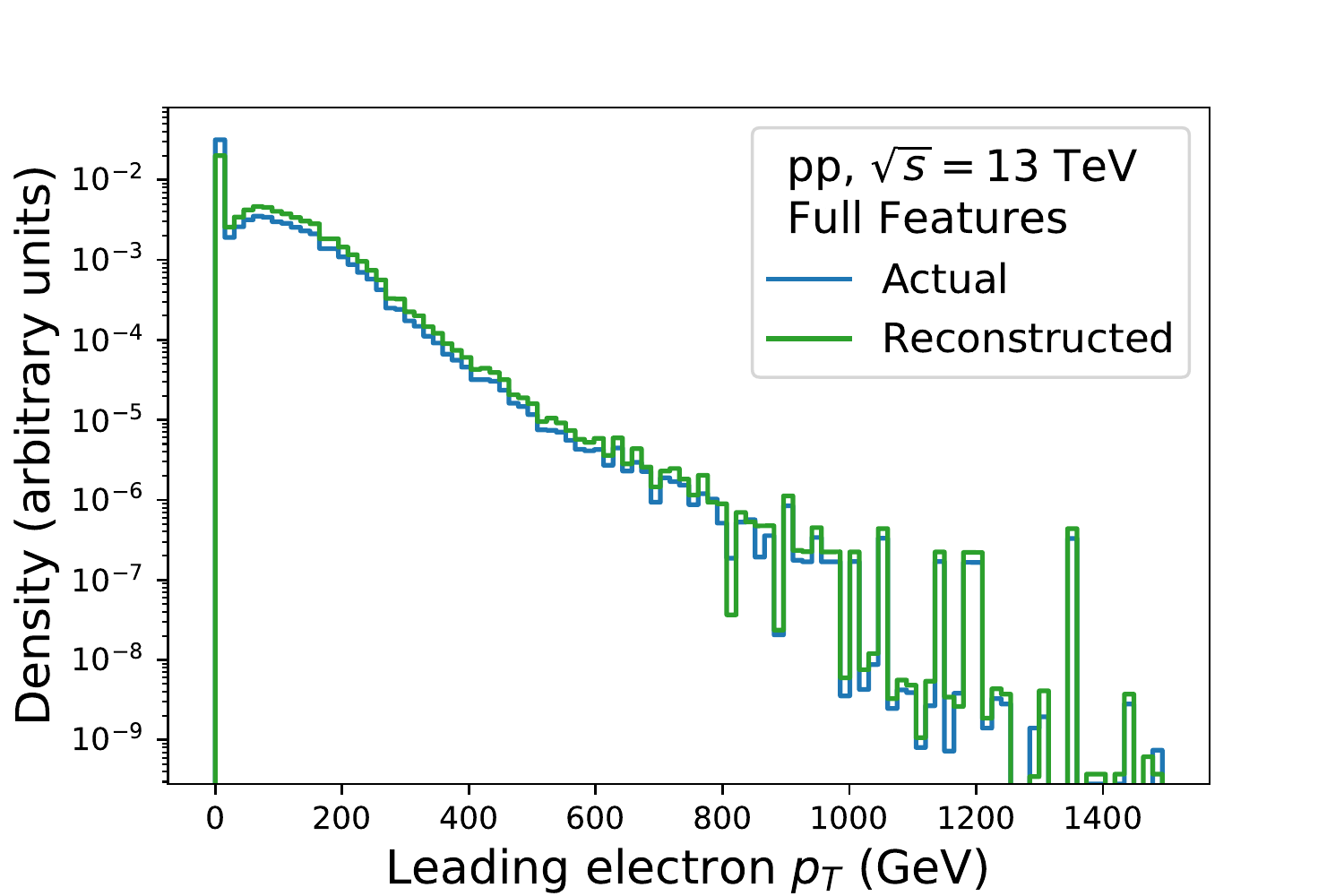}\end{subfigure}
    \begin{subfigure}[]{0.49\textwidth}\includegraphics[width=\textwidth]{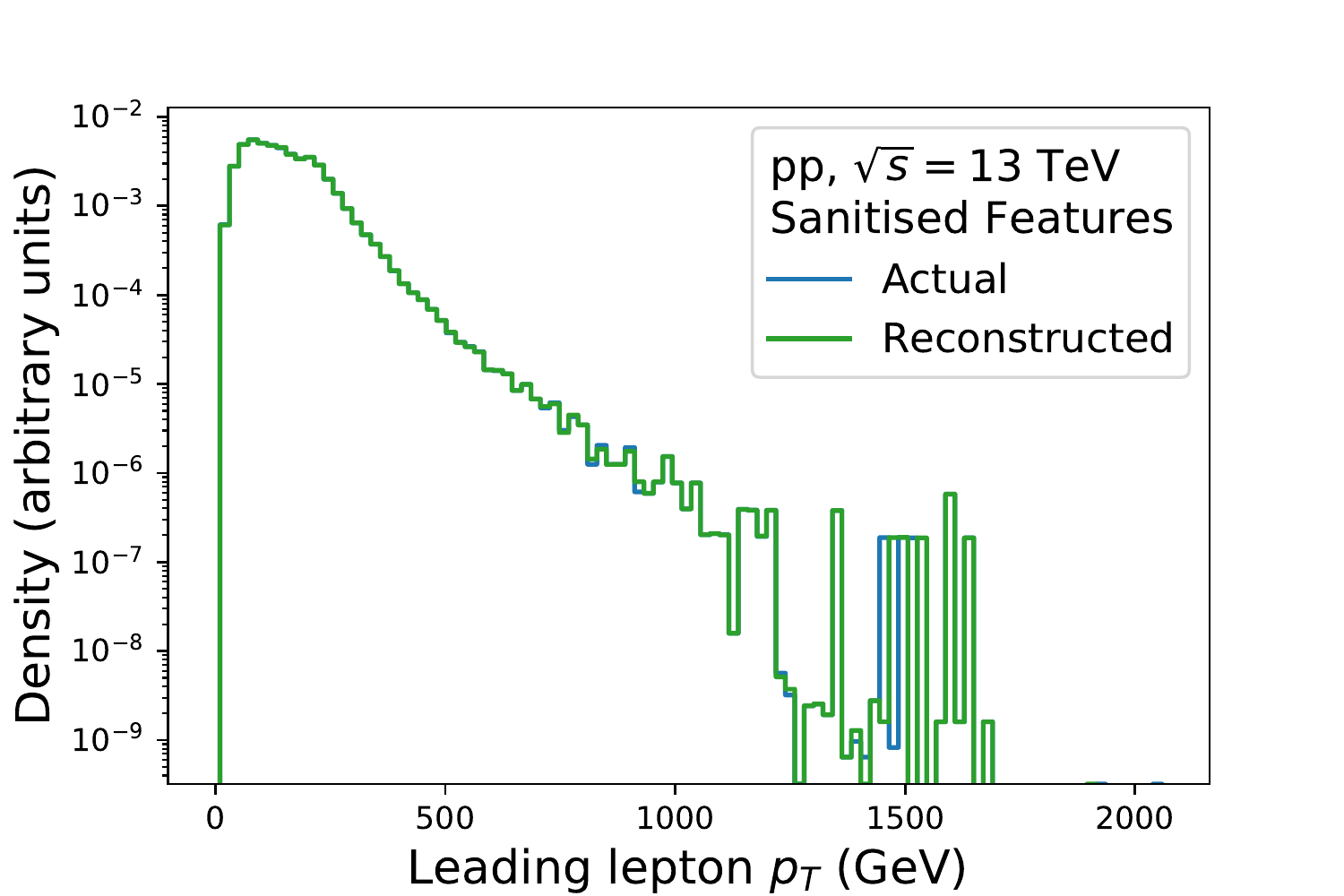}\end{subfigure}
    \caption{\label{fig:AErecon} Distribution of some of the real input features and their reconstruction by the Autoencoder on the validation set. Left: Using all features set. Right: Using sanitised features set.}
\end{figure}

% For both feature sets, full and sanitised, the AE trained with no sign of overfitting to the training data. 

{When using the full feature set, which includes events with missing reconstructed objects}, we observed that for features with pronounced accumulations at the origin, the reconstruction was degraded. In~\cref{fig:AErecon} we highlight this behaviour for three different features. For the mass of the leading large-radius jet, we notice how the accumulation in zero impacts the reconstruction of the rest of the spectrum. In the second case, concerning the $\eta$ of the leading large-radius jet, we notice that for the case with zero accumulation the AE struggles to reconstruct values away from the mean, \emph{i.e.} the origin. Removing the events without a leading large-radius jet has mitigated this problem. Finally, a similar behaviour as that of the large-radius jet is observed for the leading electron in the third case. Retaining only the two reconstructed leptons required at event pre-selection level provides a better result for the sanitised feature set.

Furthermore, as seen in the $\eta$ of the same large-radius jet distributions, removing the excess density around zero did not completely solve the reconstruction challenges of this variable. Indeed, we noticed that $\eta$ and $\phi$ variables were always difficult to reconstruct in our working methodology, {even after the hyperparameter optimisation step. We also note that transforming the 4-momenta variables to the cartesian coordinates did not resolve this issue.}
    % and even increasing the capacity of the AE did not completely solve the issue.
    {This problem highlights the challenges that DNN encounter when presented with inputs which would be better represented in varying length, such as Recurrent Neural Networks and Graph Neural Networks, which have been finding their way into HEP applications~\cite{shlomi2020graph, guest2016jet}. However, systematically study the best data representation and corresponding neural network architecture in order to provide optimal reconstruction of features using a deterministic AE is beyond the scope of this work and as such we defer such concerns to future work.}
% However, systematically study how one can improve the reconstruction of features using a deterministic AE is outside the scope of this work and as such we defer such concerns to future work.

\subsection{Anomaly scores for training and validation samples}

\begin{figure}[]
    \centering
    \includegraphics[width=0.46\textwidth]{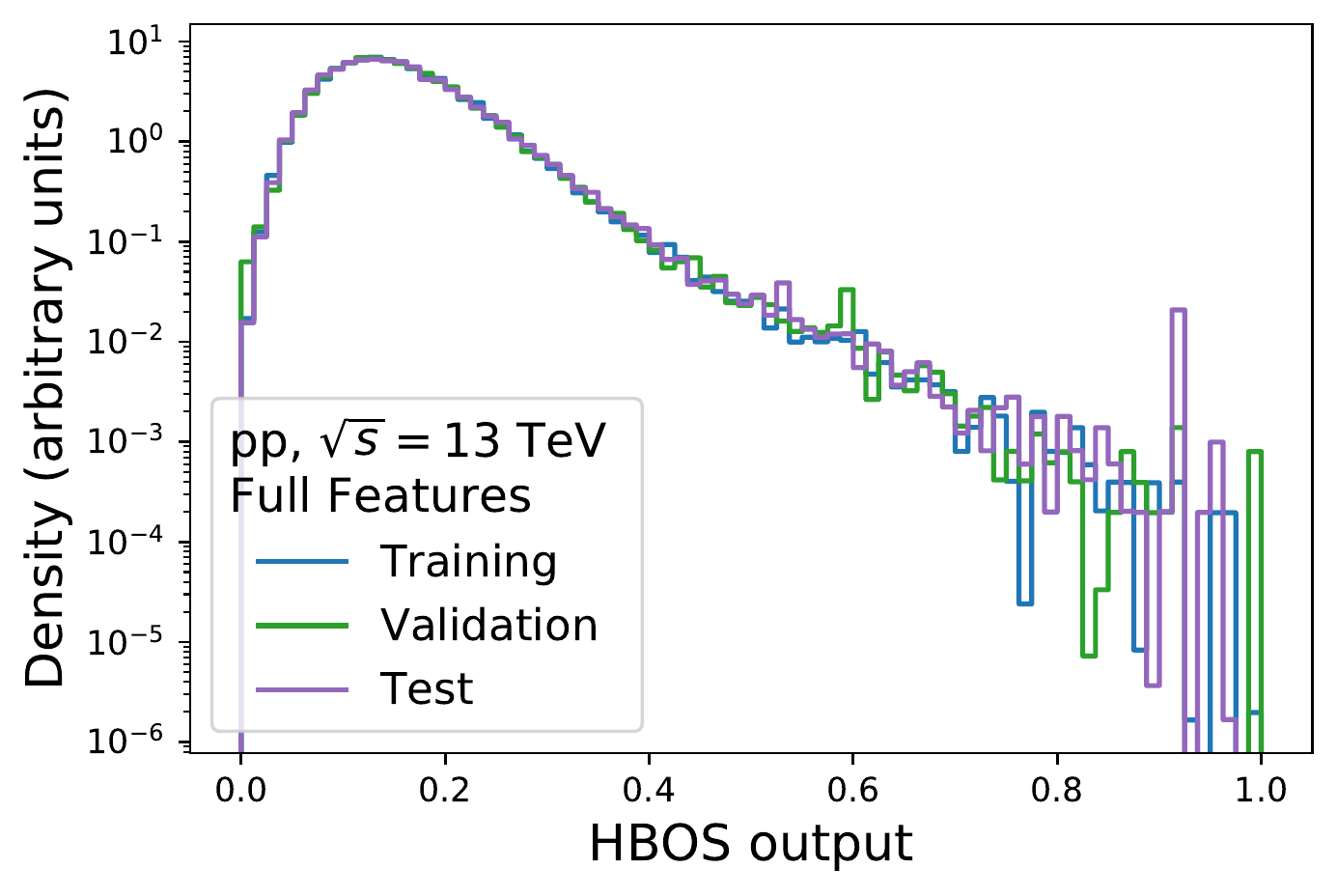} \includegraphics[width=0.46\textwidth]{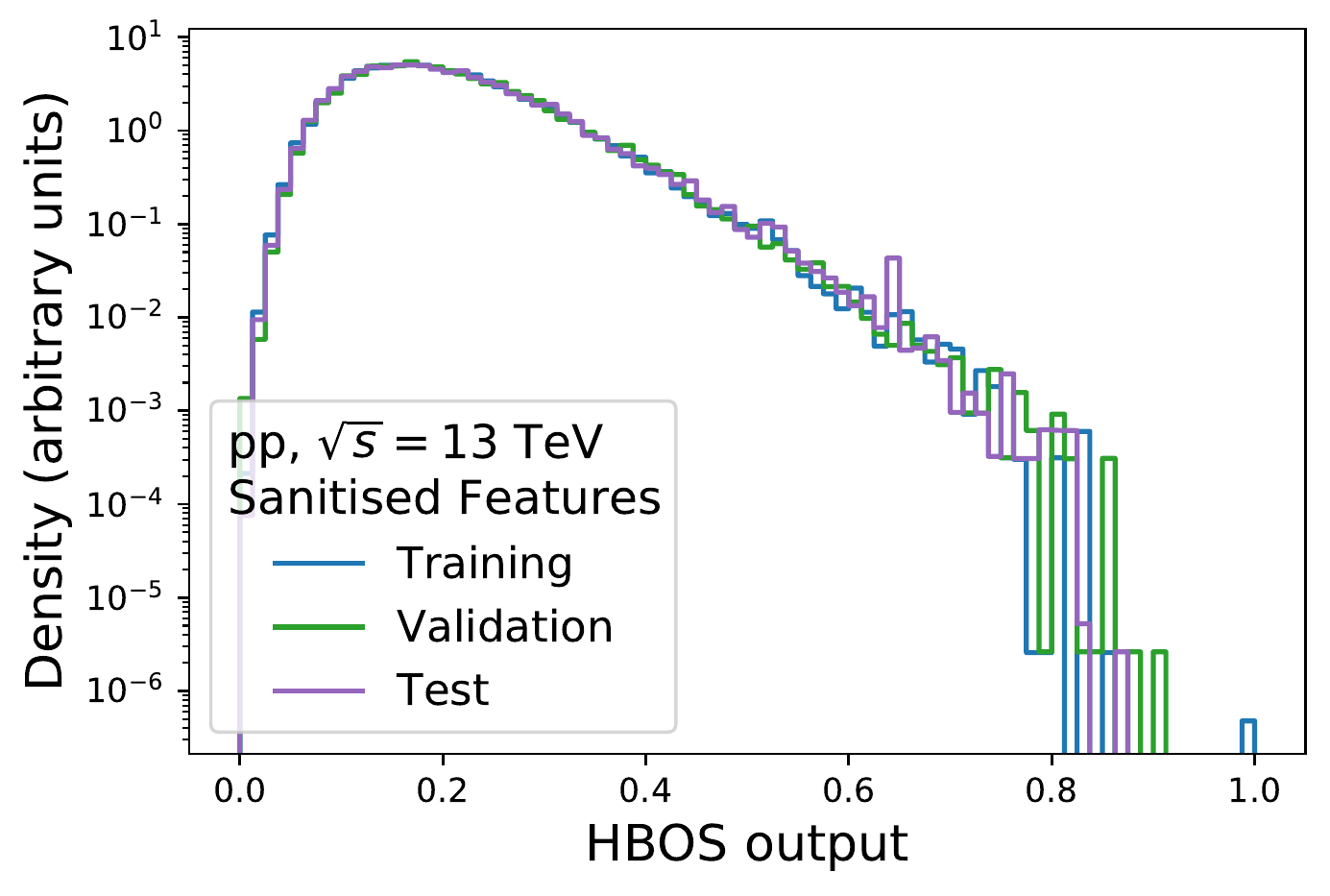}
    \includegraphics[width=0.46\textwidth]{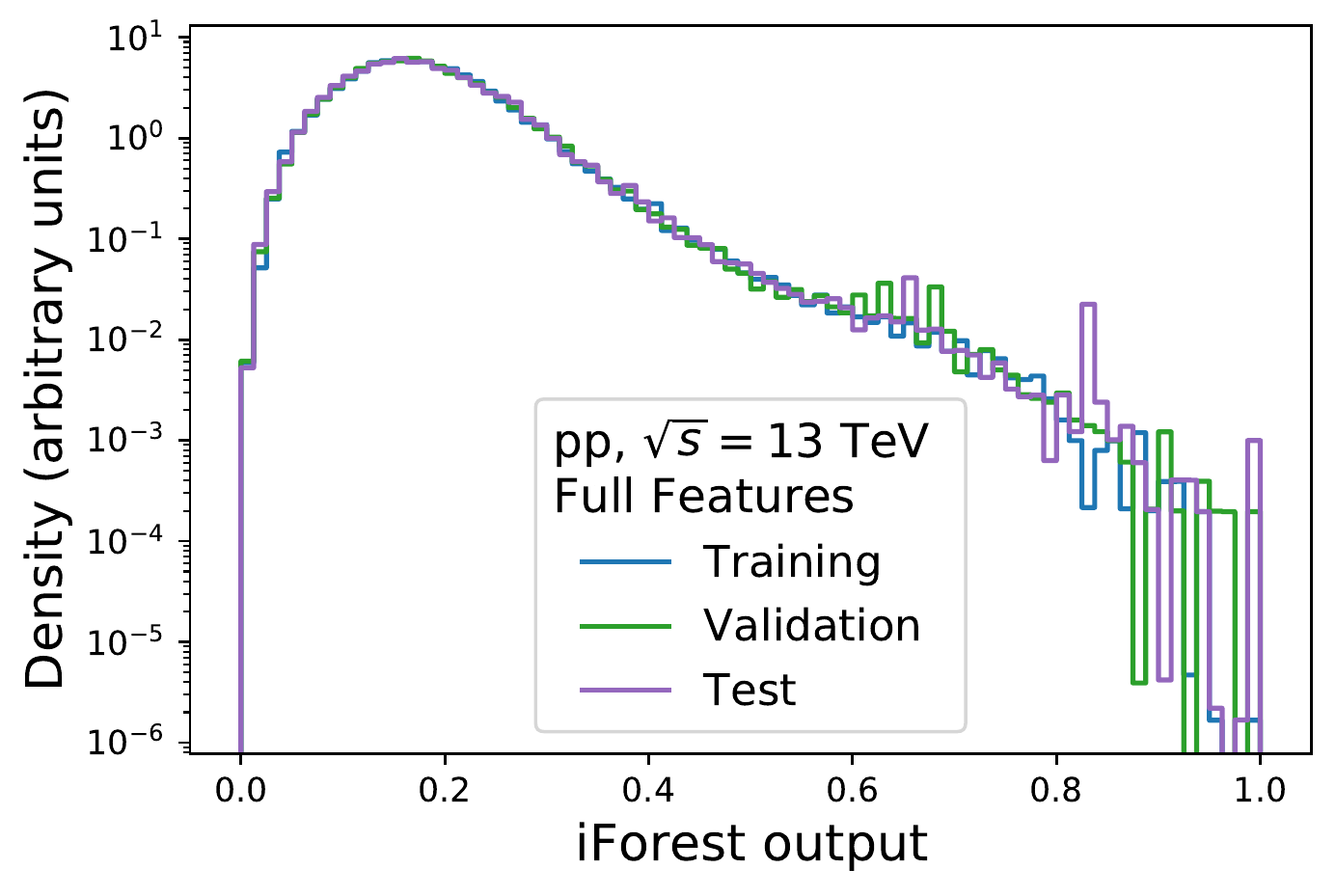} \includegraphics[width=0.46\textwidth]{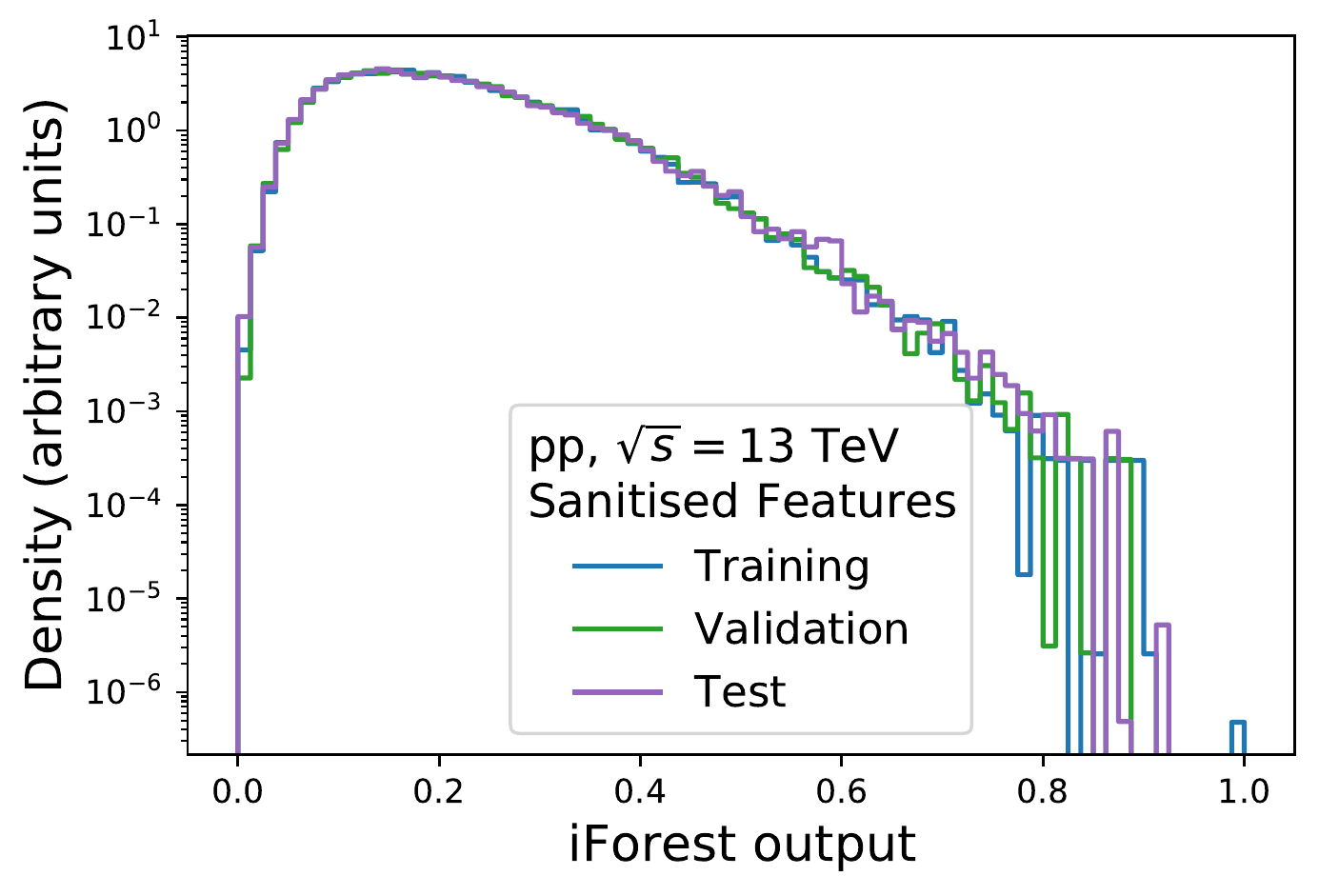}
    \includegraphics[width=0.46\textwidth]{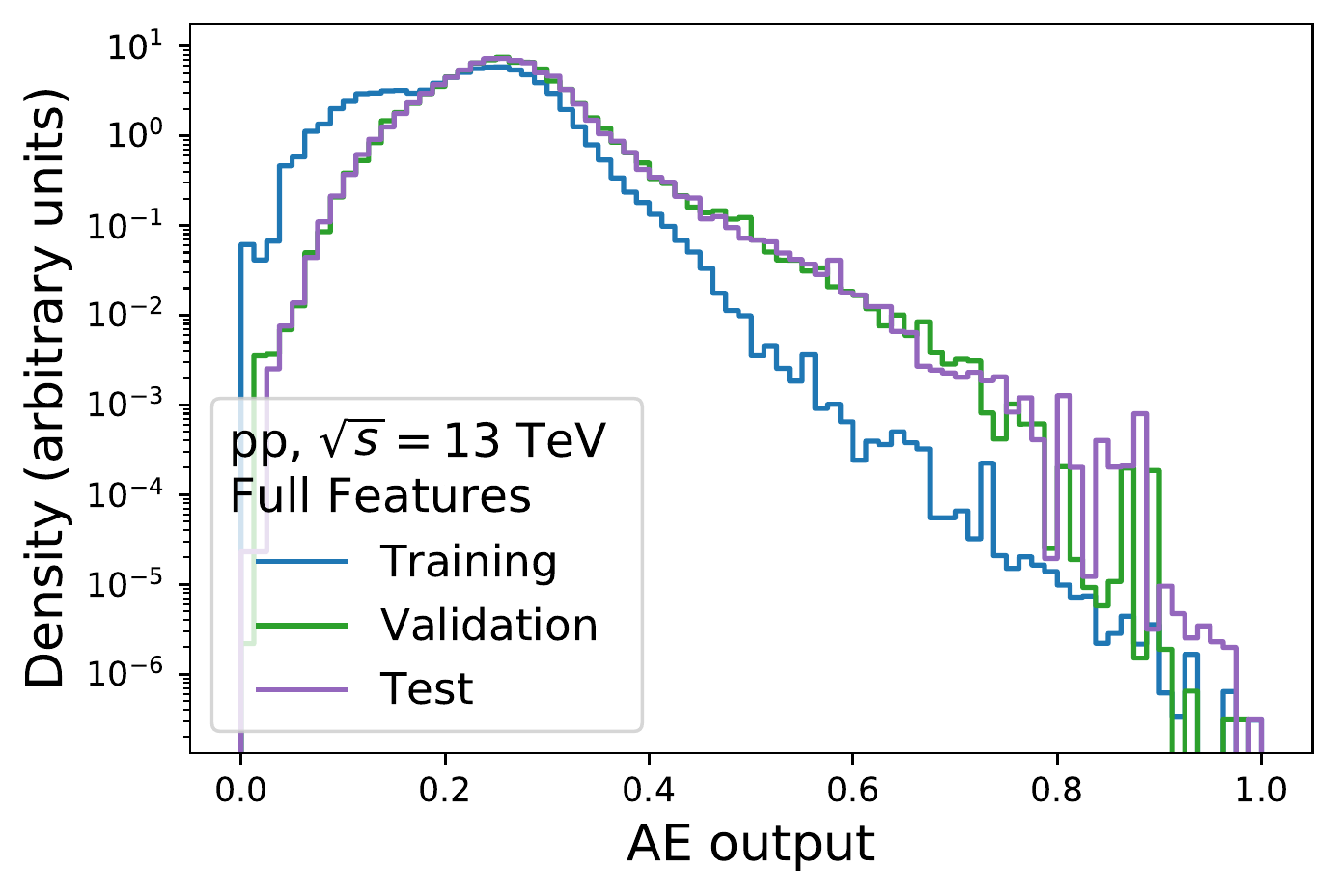} \includegraphics[width=0.46\textwidth]{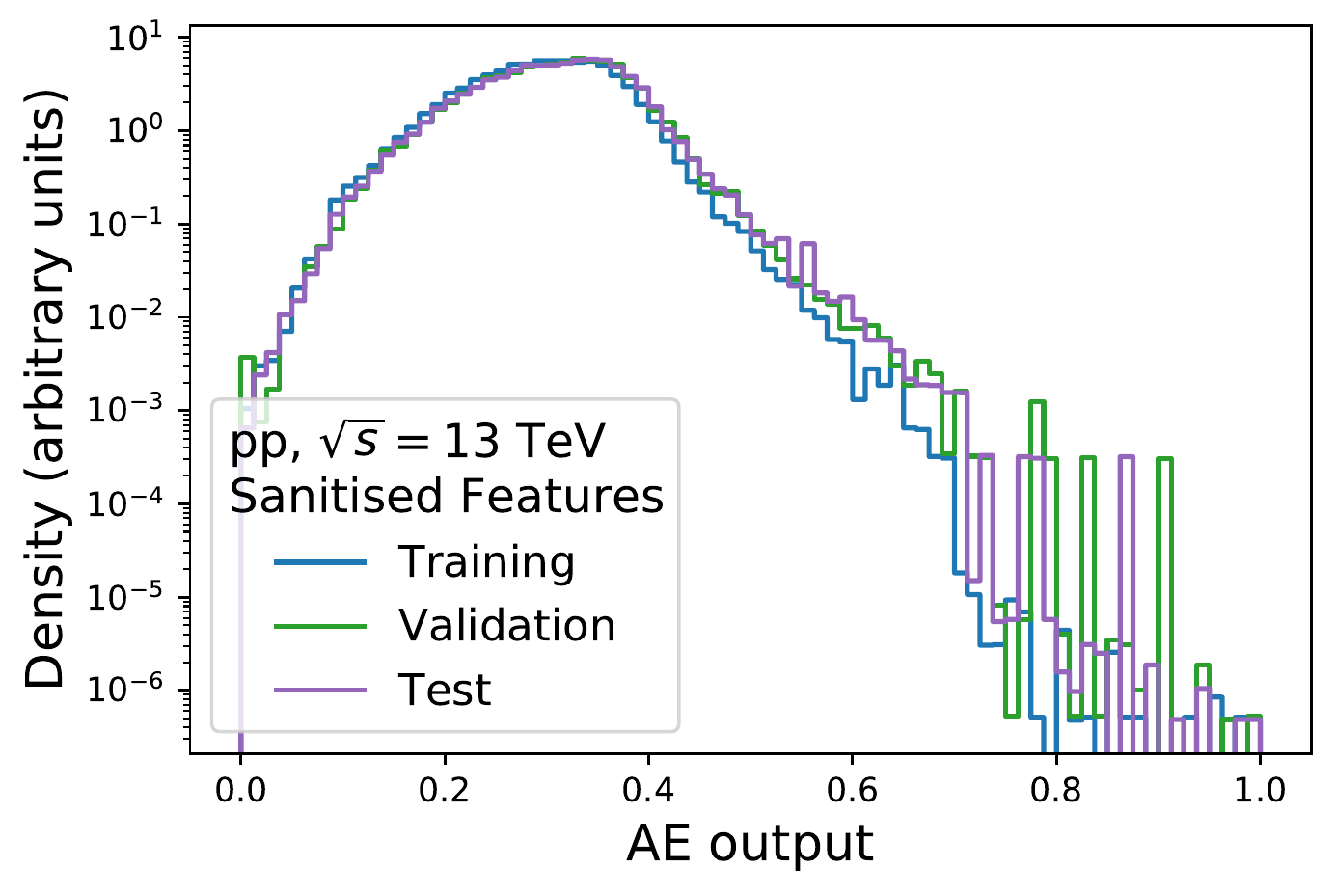}
    \includegraphics[width=0.46\textwidth]{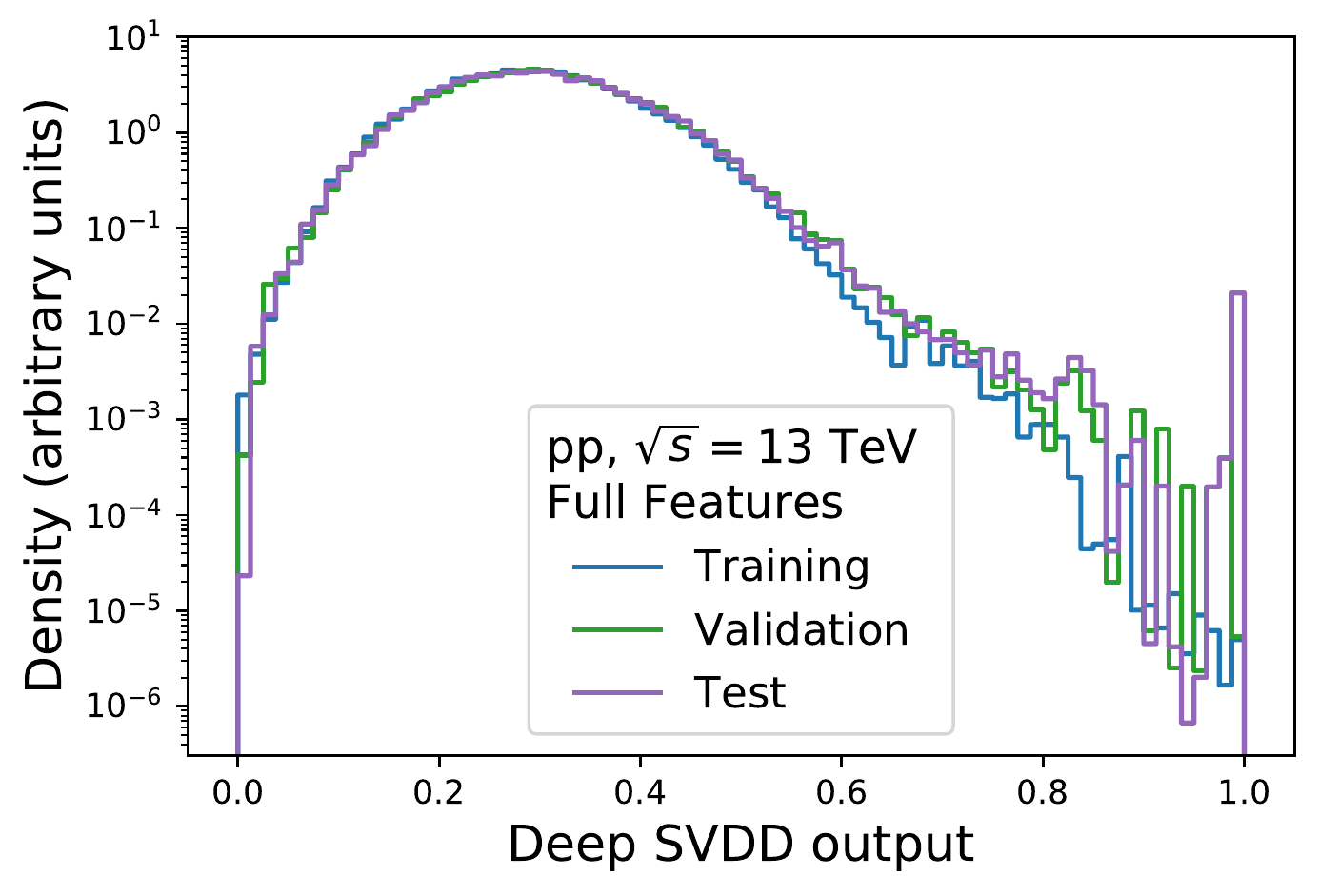} \includegraphics[width=0.46\textwidth]{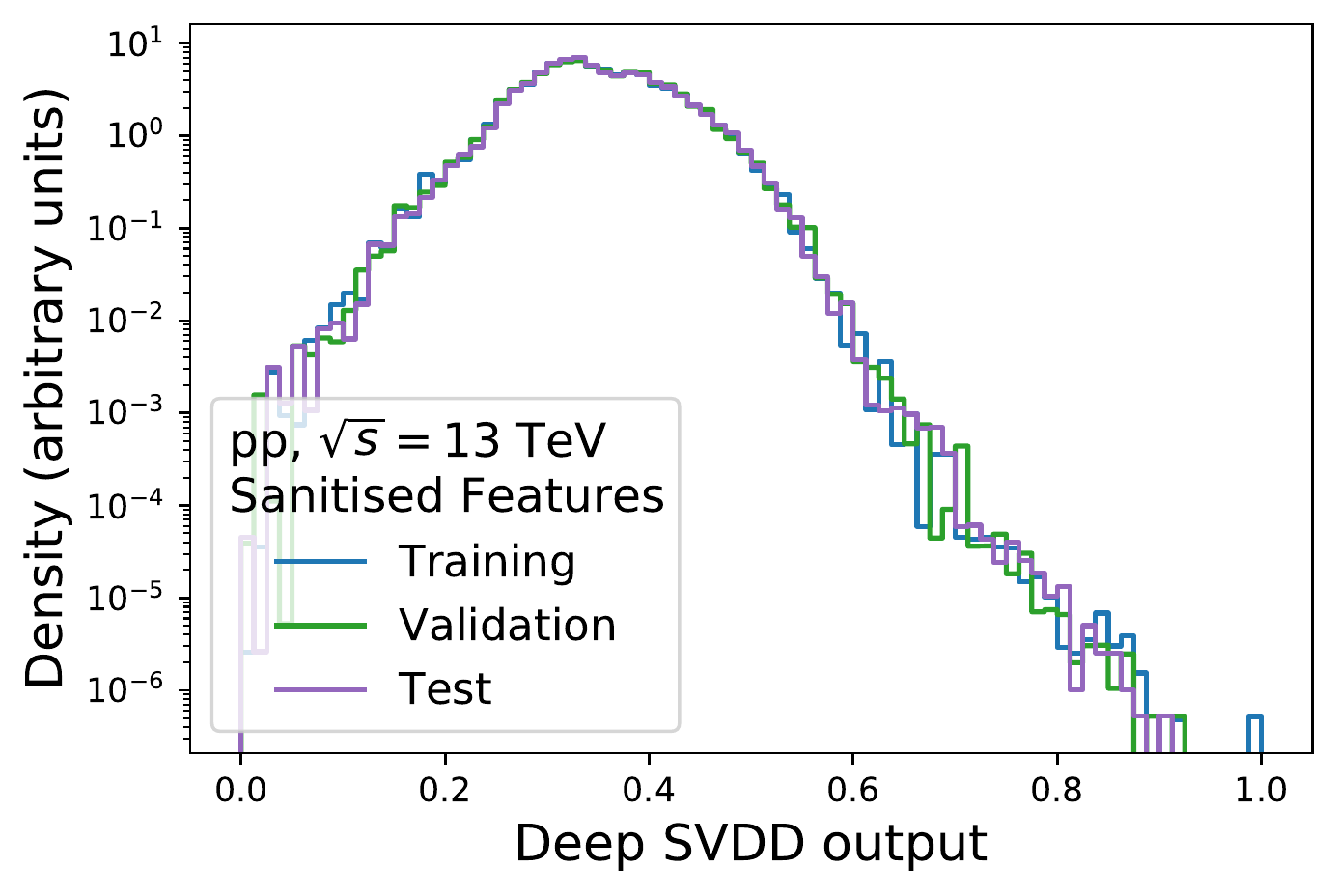}
    \caption{\label{fig:ADscore} Anomaly score for the different AD methods (HBOS, iForest, Autoencoder, Deep SVDD) for training, validation, and test data. The distributions are normalised to the unit area. Left: Using all features set. Right: Using sanitised features set.}
\end{figure}

The anomaly score distributions for each of the four AD methods are presented in~\cref{fig:ADscore} for the {training, validation, and test samples and both feature sets. We notice that for the full features set both the deep AD models manifest a more pronounced difference between the training and validation sample distributions, with a significant difference in the AE case. However, we also observe that the validation sample follows same distribution as the test sample, where the upper limits on signal strength, i.e. the physical application, will be calculated. This provides some confidence that, even though these methods are overfitting to the training data, the observed behaviour for the validation set is expected to carry to the test set.}

\begin{figure}[]
    \begin{subfigure}[]{0.49\textwidth}\includegraphics[width=0.9\textwidth]{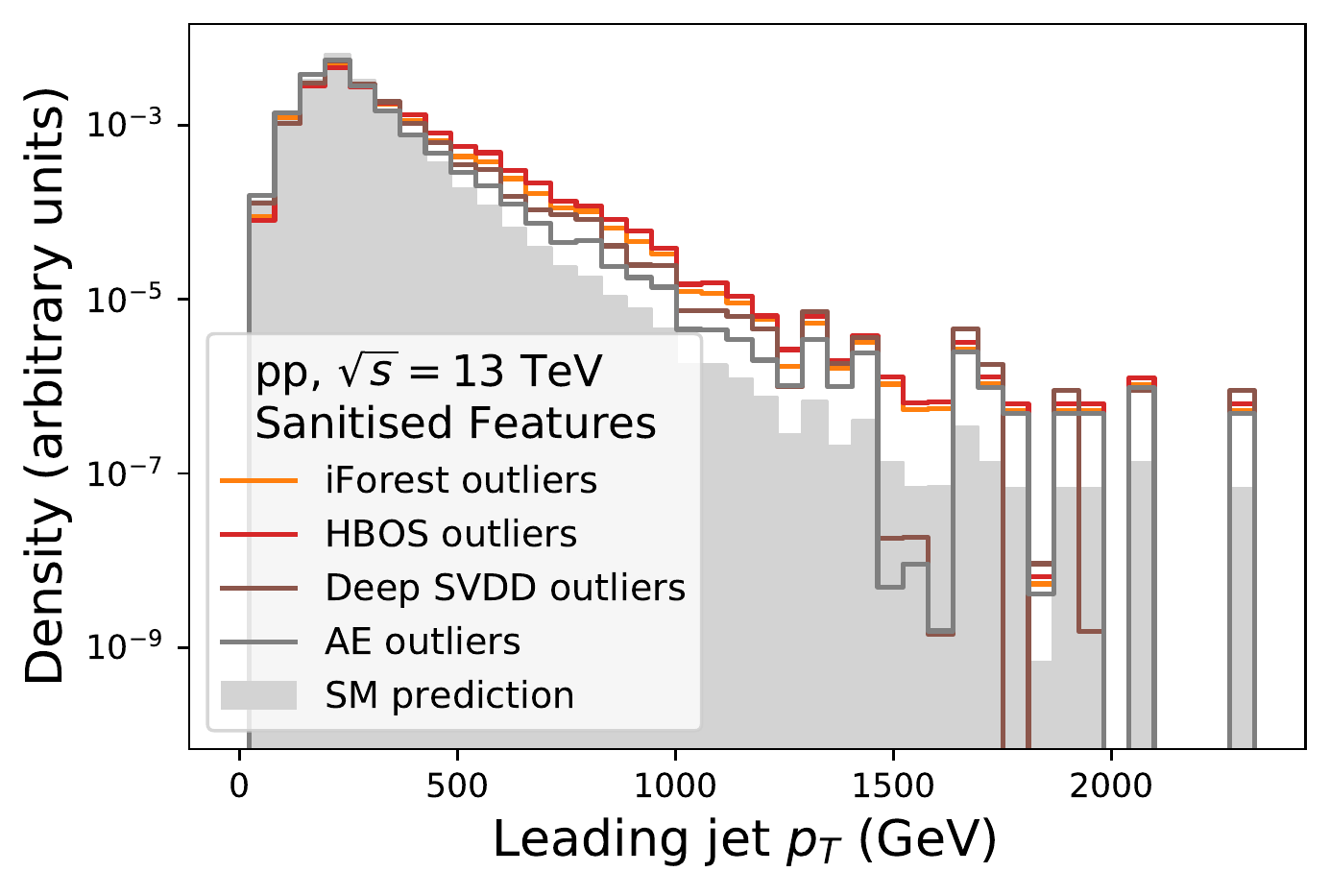}\end{subfigure}
    \begin{subfigure}[]{0.49\textwidth}\includegraphics[width=0.9\textwidth]{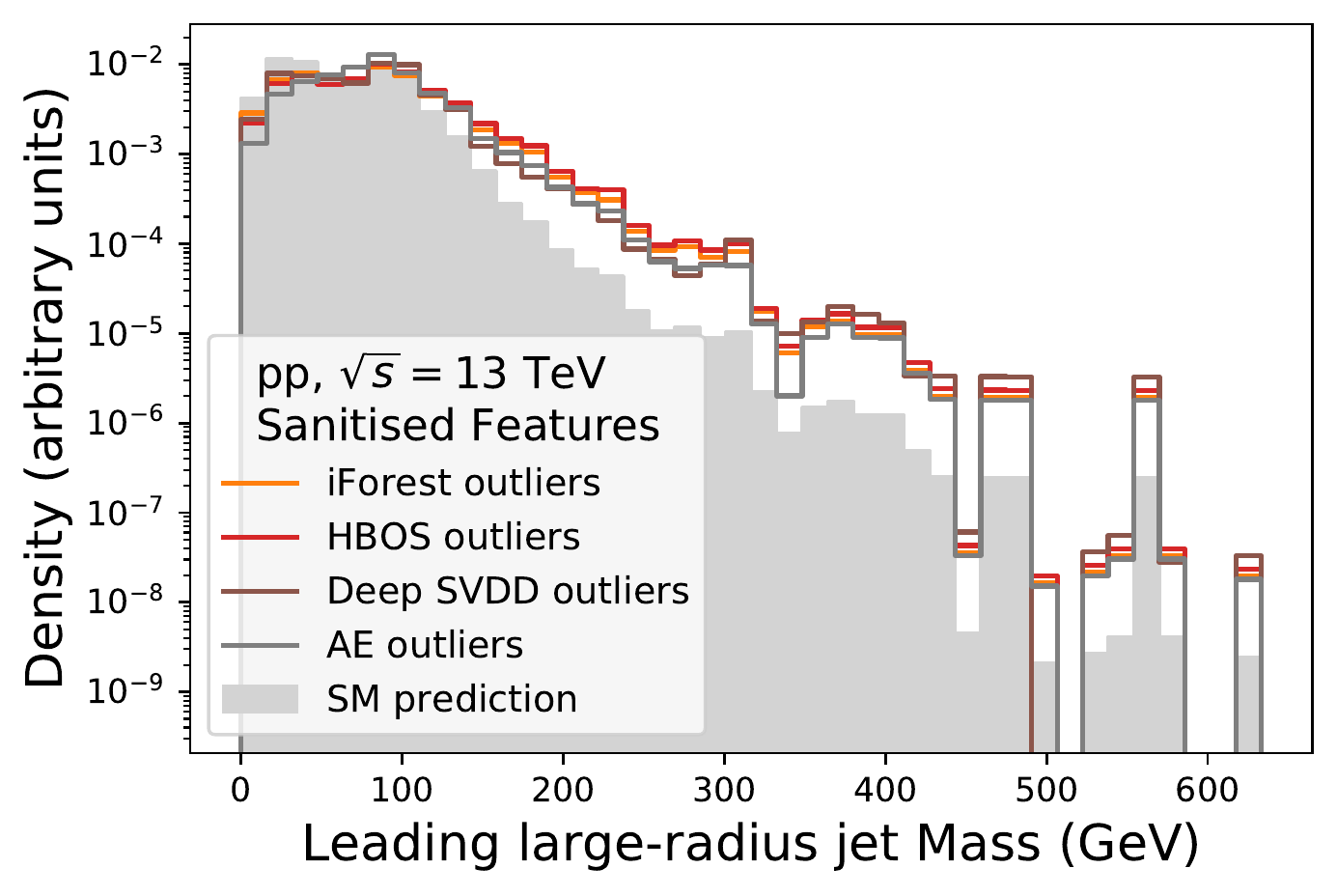}\end{subfigure}\\
    \begin{subfigure}[]{0.49\textwidth}\includegraphics[width=0.9\textwidth]{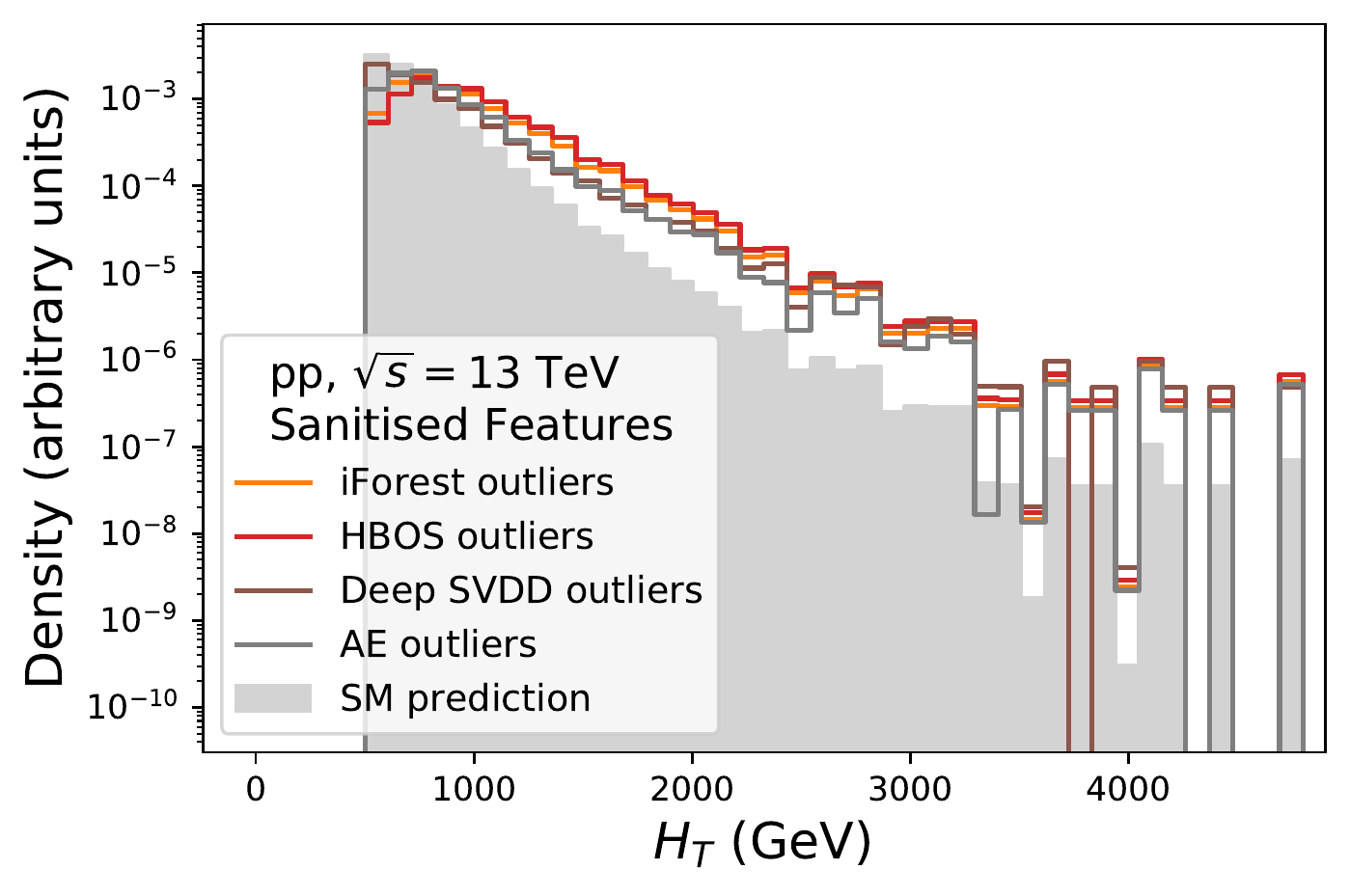}\end{subfigure}
    \begin{subfigure}[]{0.49\textwidth}\hspace*{.6mm}\includegraphics[width=0.893\textwidth]{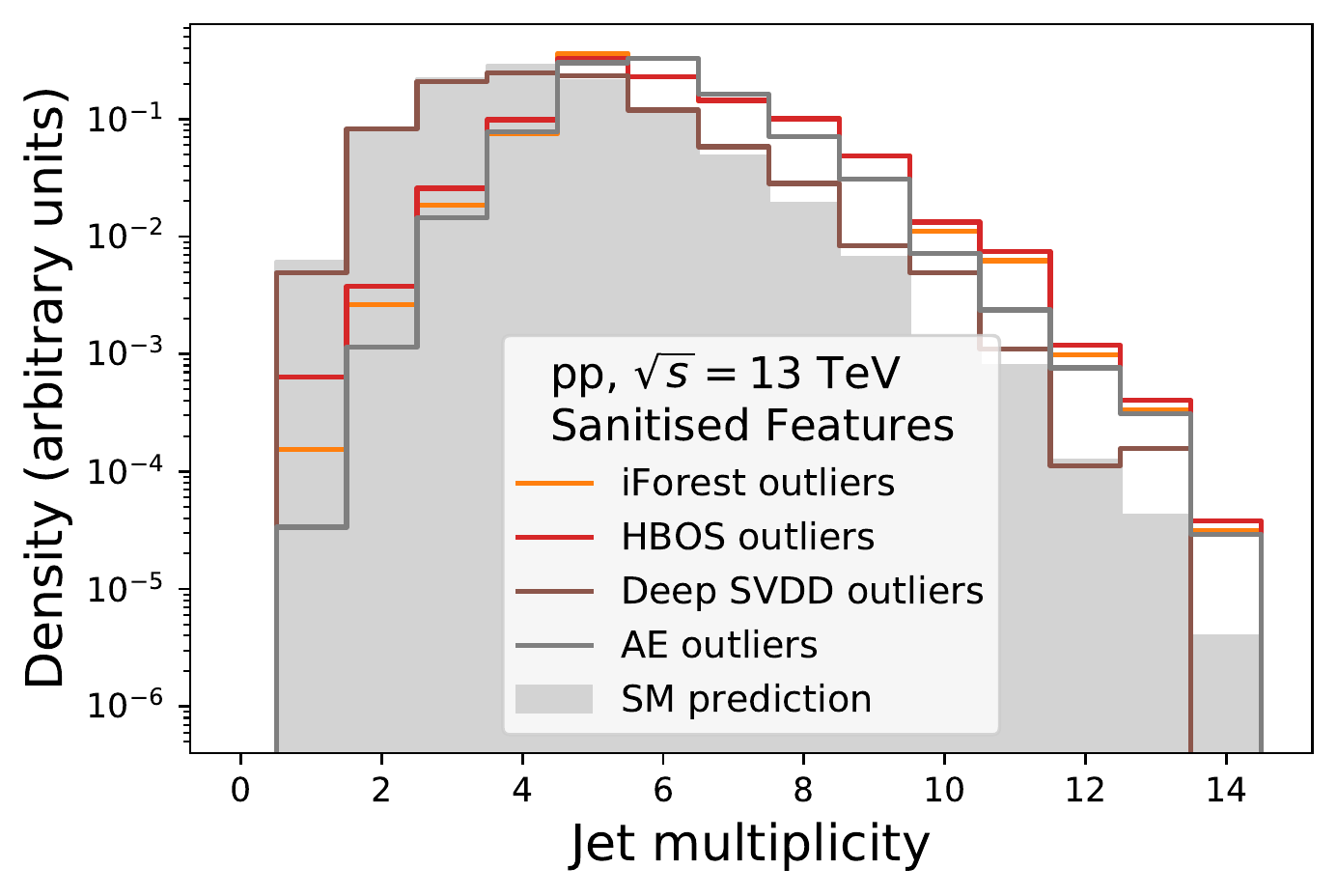}\end{subfigure}
    \caption{\label{fig:quantile} Distribution of some of the input features for the full validation set and for the 10\% outlier quantile according to the anomaly score for the different AD methods using sanitised features. All distributions are normalised to the unit area.}
\end{figure}

In~\cref{fig:quantile} we show the distributions of four example features for the 10\% most anomalous events under each AD method score -- \emph{i.e.} the events whose score lies in the 10\% outlier quantile calculated on the validation distribution shown in~\cref{fig:ADscore}, using the sanitised feature set. The figure shows that the AD algorithms are capturing the tails of distributions. However, we can see from the Jet Multiplicity distribution that the Deep SVDD seems to be capturing different events than the remaining AD methods, manifesting that the anomaly/outlyingness of an event can be very much dependent of the type of AD algorithm.

\begin{figure}[]
    \centering
    \includegraphics[width=0.83\textwidth]{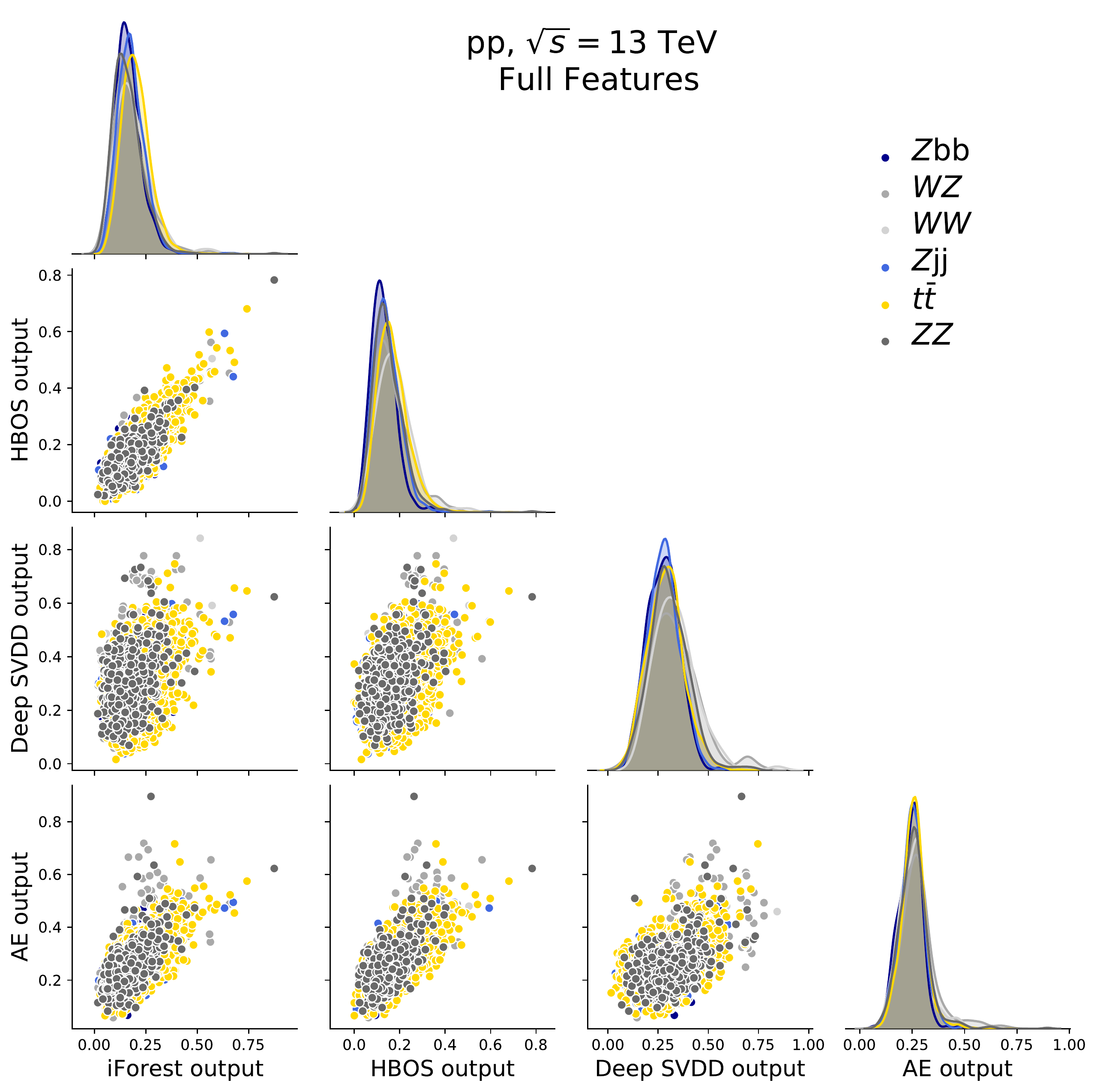}
    \caption{\label{fig:scatter_plots} Two-dimensional distribution of the anomaly scores for the different AD methods per SM process -- $t\bar t$, $Z$+jets and diboson -- using all features set. Diagonal: Distribution of the anomaly score per SM process.}
\end{figure}

\begin{figure}[]
    \centering
    \includegraphics[width=0.83\textwidth]{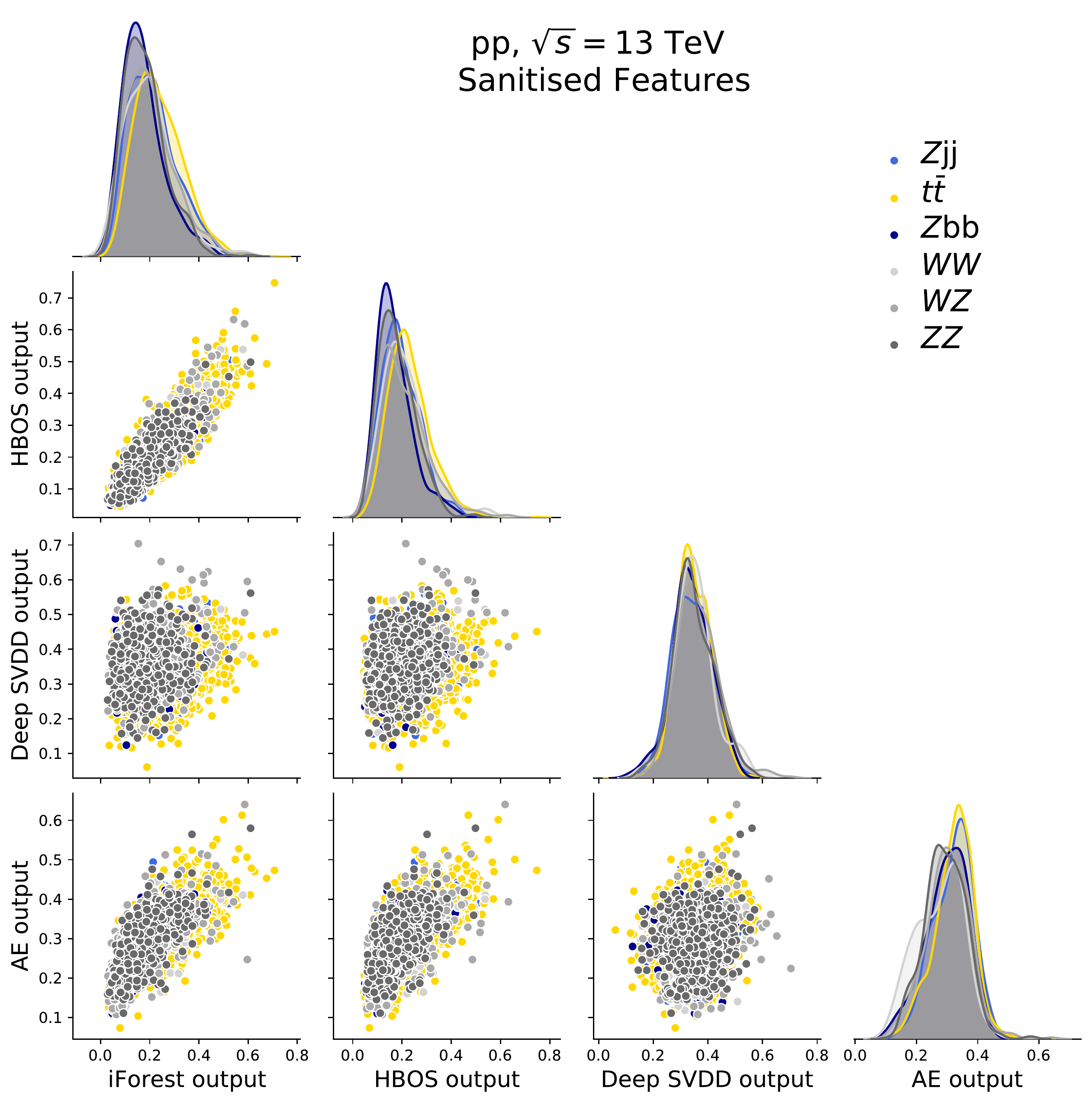}
    \caption{\label{fig:scatter_plots_nozeros} Two-dimensional distribution of the anomaly scores for the different AD methods per SM process -- $t\bar t$, $Z$+jets and diboson -- using sanitised features set. Diagonal: Distribution of the anomaly score per SM process.}
\end{figure}

In~\cref{fig:scatter_plots,fig:scatter_plots_nozeros} we present the distributions and the scatter plots of the anomaly scores for each process of the SM cocktail used in the AD model training. {The correlation trends are similar across the individual SM processes.} We notice how the shallow methods are highly related between each other for both feature sets {over the validation sample}. In contrast, both deep models show looser relation between their predictions and the shallow predictions, and amongst themselves, for both feature sets. More interestingly, we notice how the Deep SVDD and the AE have a small correlation in the sanitised set. Again, these results point to the fact that different AD algorithms will be capturing different anomalous events.

\section{Comparison of the AD methods for benchmark signals}

In this section, we assess the performance of the trained AD models to discriminate signals from new physics, not present in the SM cocktail used for their development. The performance metric is based on the 95\% confidence level (CL) upper limit on the signal strength $\mu$, defined as the ratio between the expected upper limit on the signal cross-section, normalised to the corresponding theory prediction, computed at leading order. Such limits were obtained by fitting the AD score distribution of the test data set and were computed using the $CL_s$ method~\cite{Read:2002hq}, as implemented in OpTHyLiC~\cite{opthylic}. Poissonian statistical uncertainties on each bin of the distributions were included in the limit computation, assuming an integrated luminosity of 150~fb$^{-1}$.

\subsection{Anomaly score distributions}

\begin{figure}[]
    \begin{center}
        \includegraphics[width=0.44\textwidth]{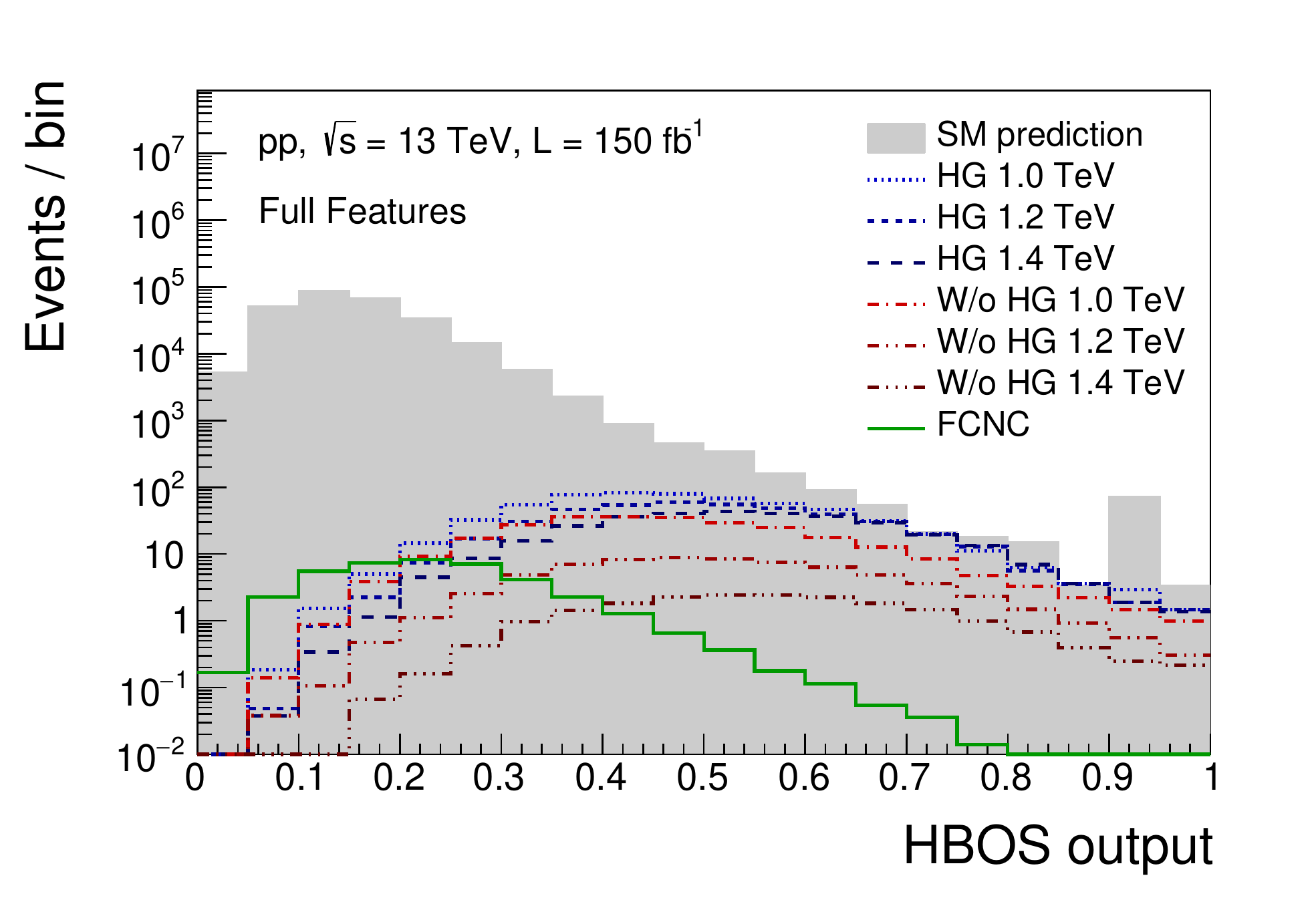} \includegraphics[width=0.44\textwidth]{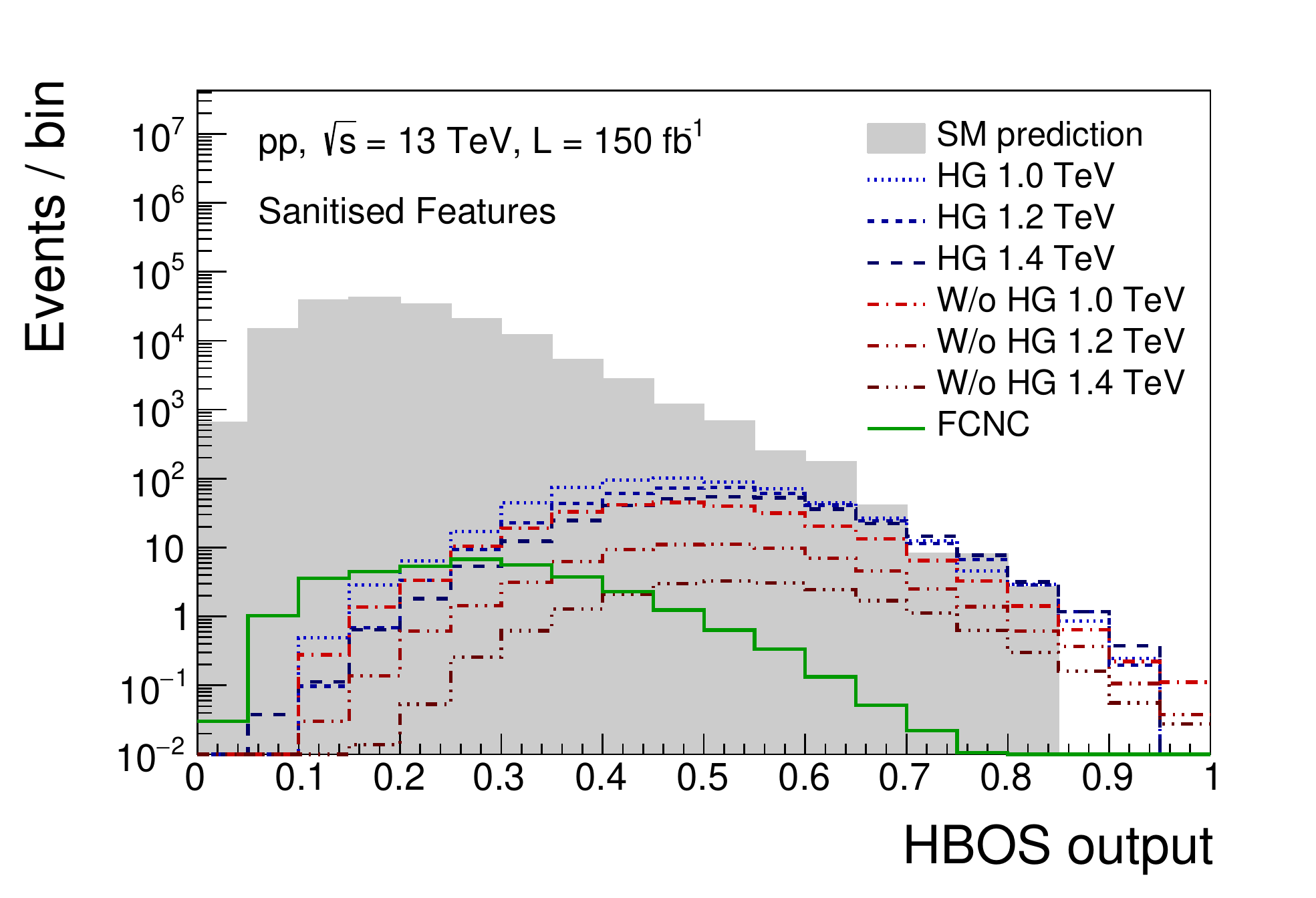}
        \includegraphics[width=0.44\textwidth]{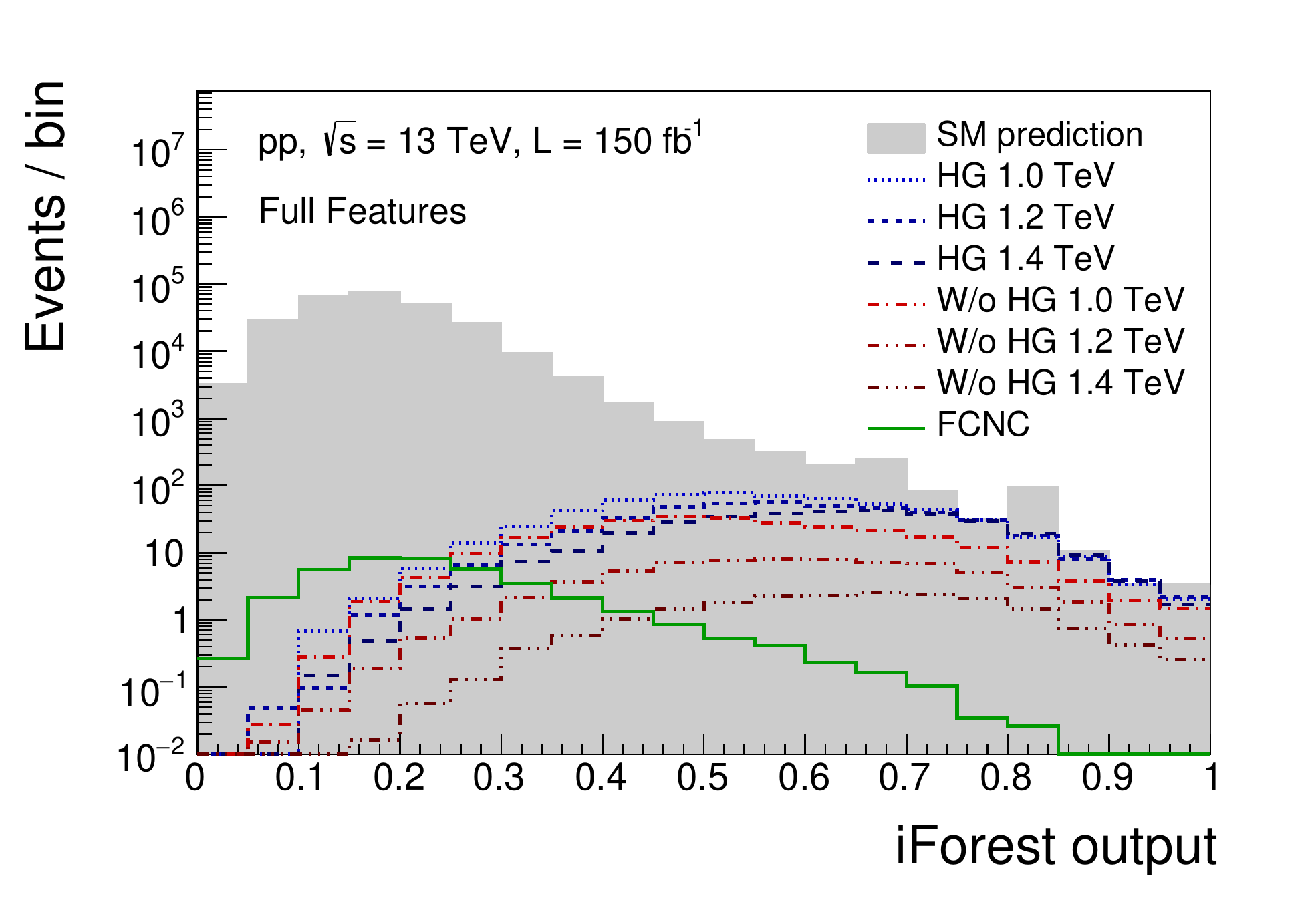} \includegraphics[width=0.44\textwidth]{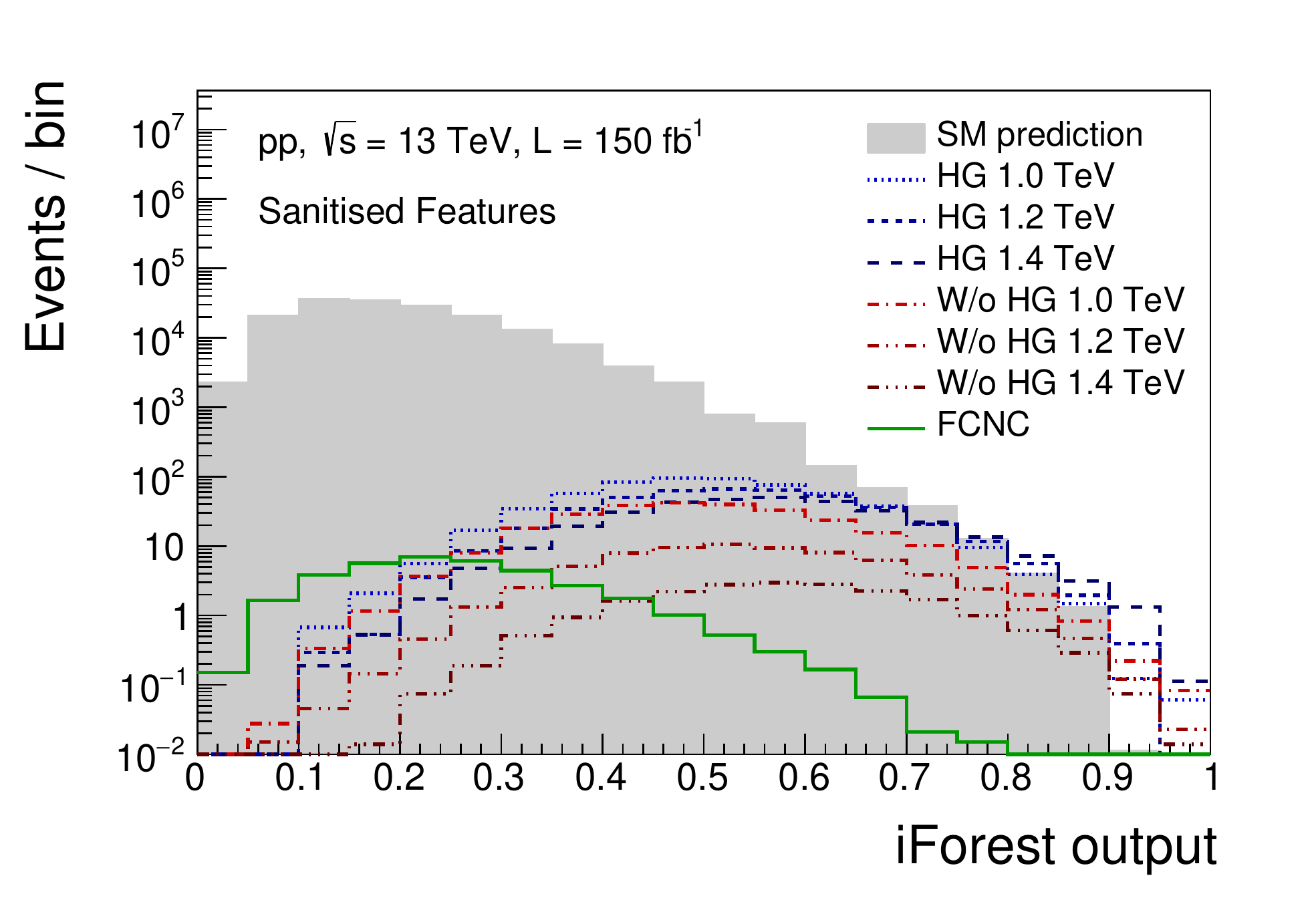}
        \includegraphics[width=0.44\textwidth]{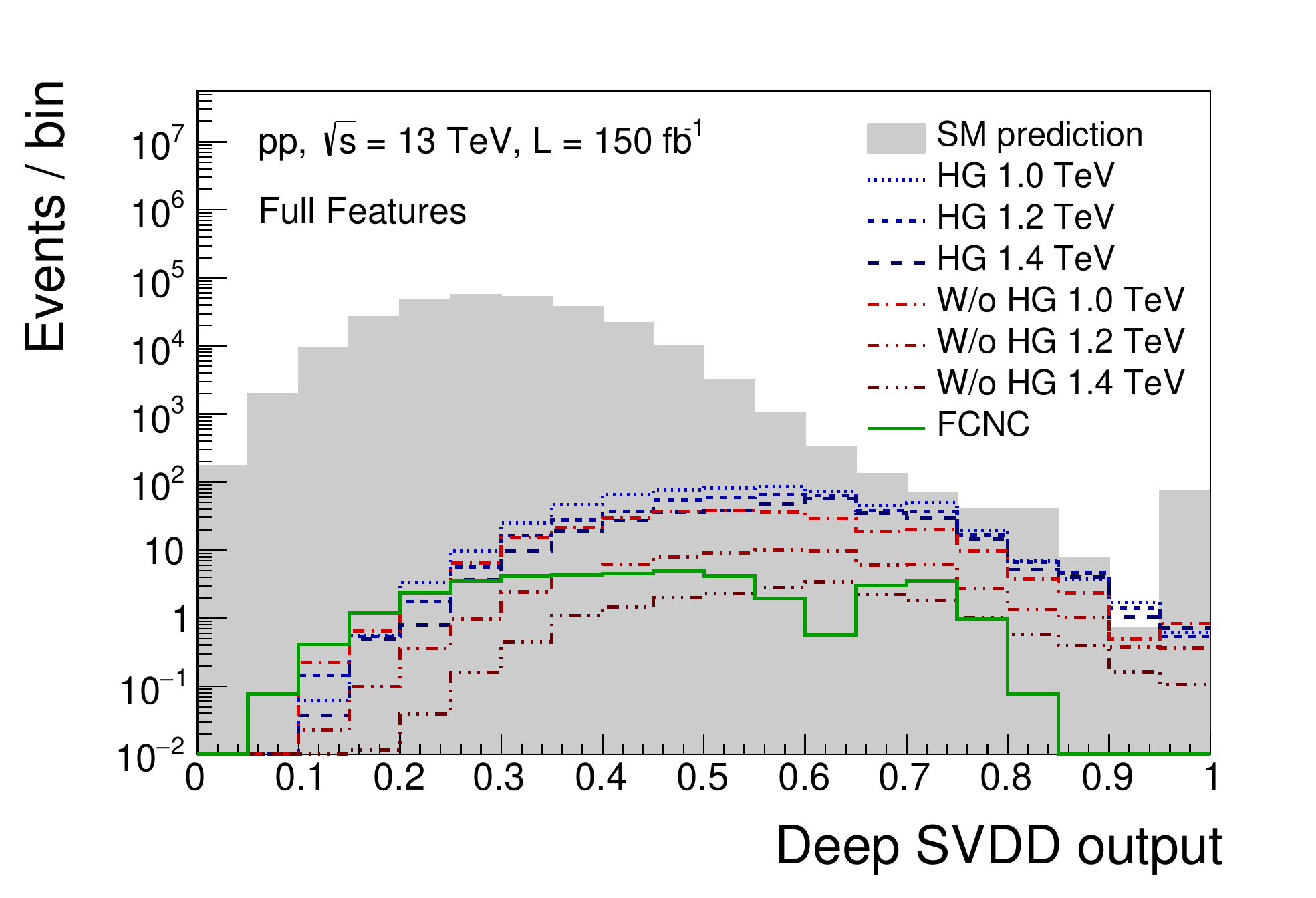} \includegraphics[width=0.44\textwidth]{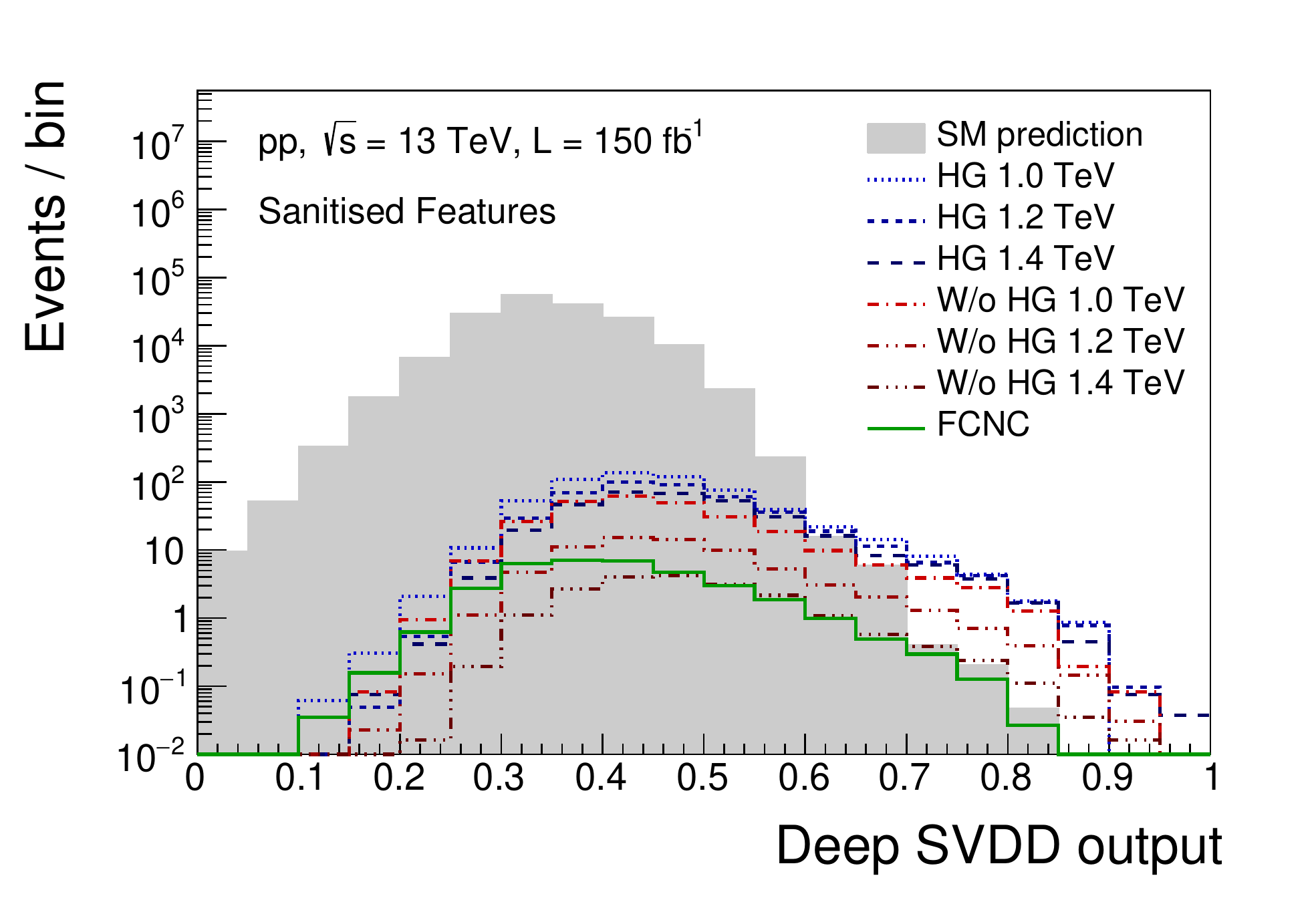}
        \includegraphics[width=0.44\textwidth]{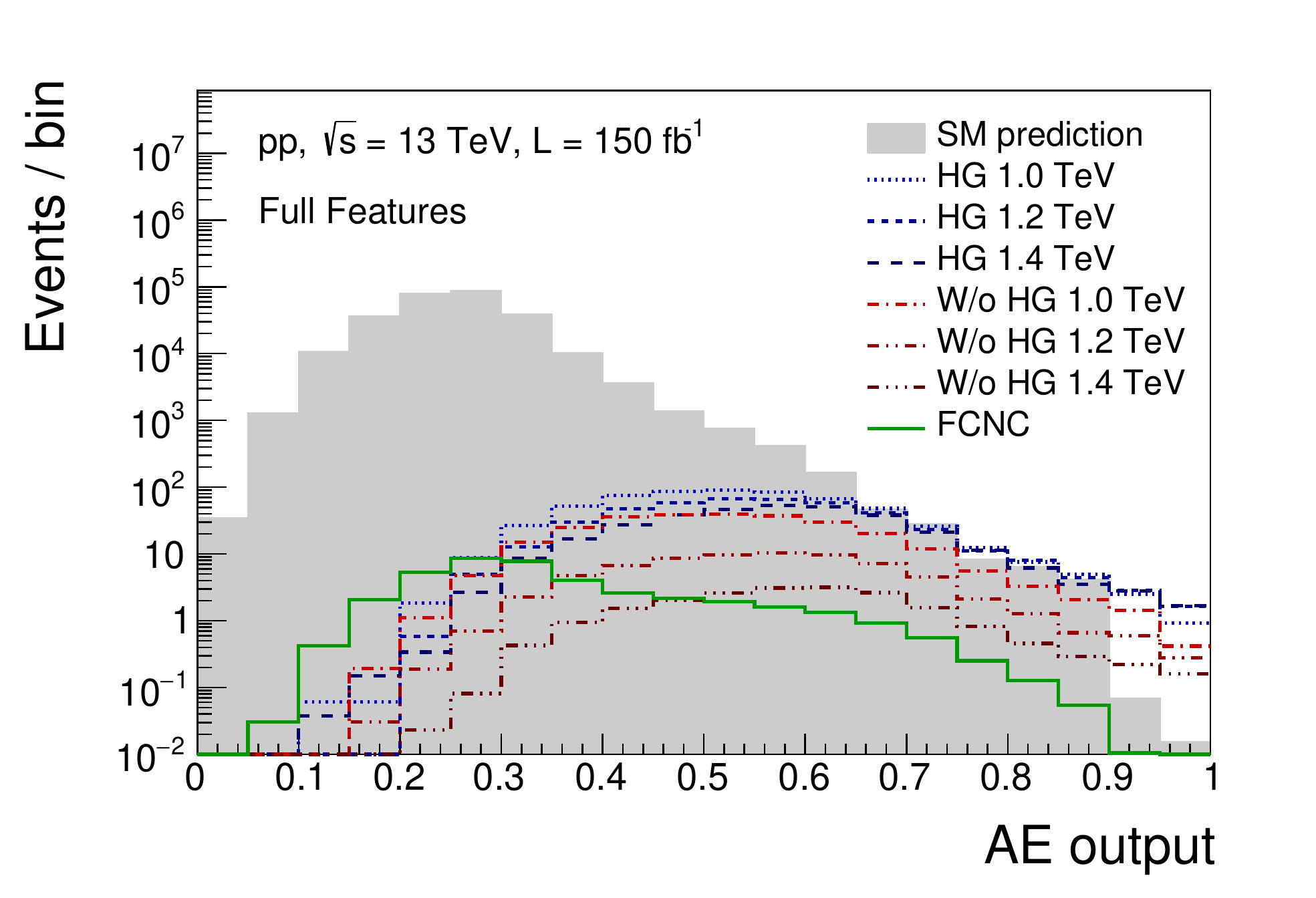} \includegraphics[width=0.44\textwidth]{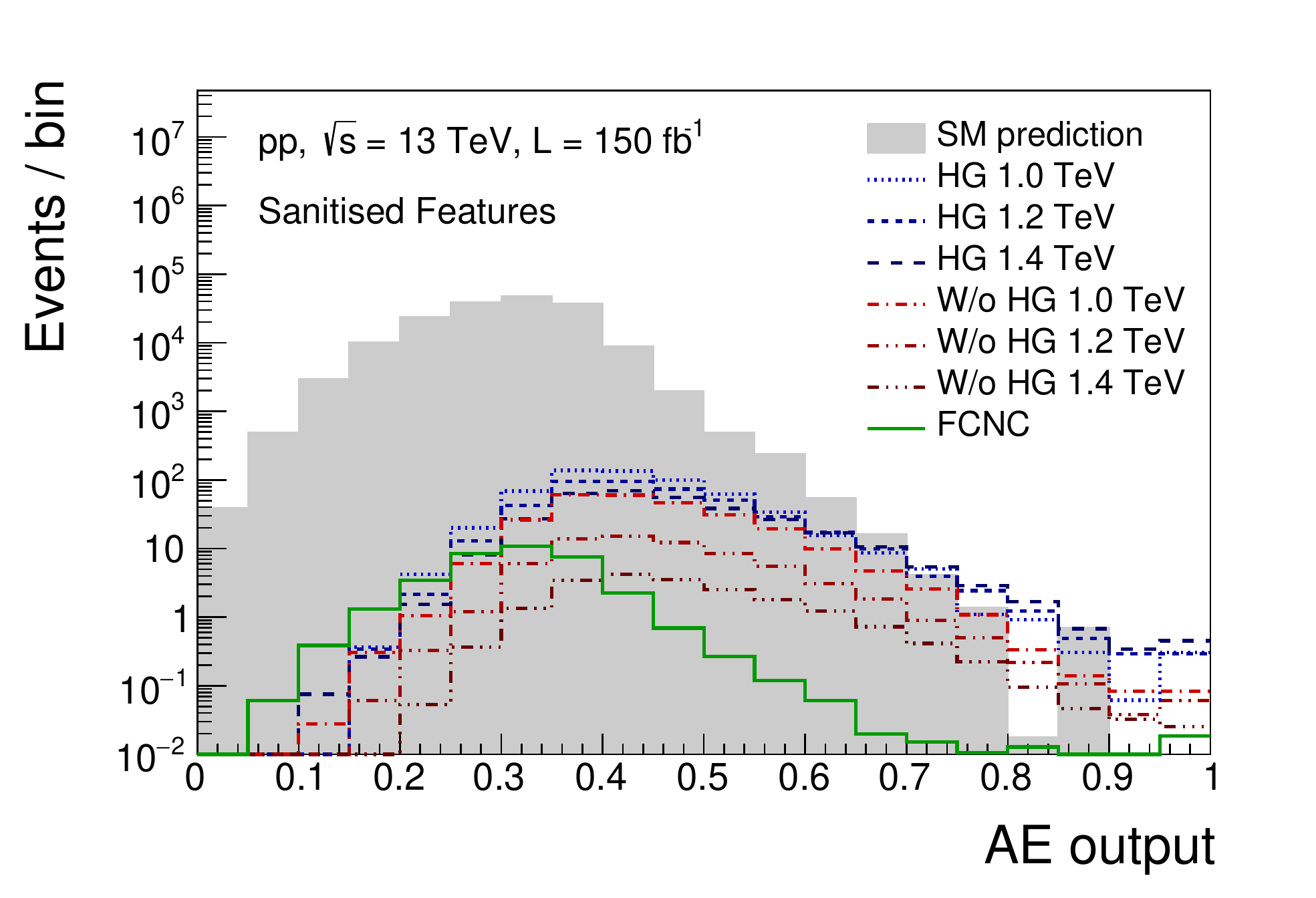}
        \caption{\label{fig:fit_inputs} Distribution of the AD discriminant for the SM prediction and each signal type: $tZ$ production by FCNC, $T\bar T$ production via heavy gluon or without heavy gluon for $m_T=\{1.0, 1.2, 1.4\}$~TeV. The distributions are normalised to the generation cross-section and to an integrated luminosity of 150~fb$^{-1}$. Left: Using all features set. Right: Using sanitised features set.}
    \end{center}
\end{figure}

In~\cref{fig:fit_inputs} we present the output distributions of the four AD models trained on both feature sets, for the SM prediction and each benchmark signal. We observe that the shallow methods have similar behaviour for both feature sets, and in each of them, the FCNC signal follows a distribution that is very close to the one followed by the SM processes. In contrast, the vector-like $T$-quarks are being assigned on average higher anomaly scores.

For the deep models, we observe a significant difference in distribution 
shapes when we switch from the full feature set to the sanitised feature 
set. In particular, we notice how the Deep SVDD provides significant better 
capacity to isolate signal with the sanitised feature set. For the AE, the 
FCNC distribution becomes more similar to the SM background, as it happens to 
the shallow methods. In both cases, the anomaly score distributions for the 
signals have their mass shifted to the right, meaning that on average 
abnormal signals have higher anomaly scores than the SM events and that this 
behaviour is more noticeable in the deep models.

\subsection{Expected upper limits}

\begin{table}[h]
    \scriptsize
    \caption{\label{tab:limits} 95\% CL upper limit on the signal strength $\mu$ of each benchmark signal for the different AD methods using the full feature set and the sanitised set and for a dedicated supervised DNN model trained on the full feature set. The statistical uncertainties, including the effect from limited statistics in the simulated datasets, are also shown.}
    \begin{center}
        \hspace{-0.6cm}
        \begin{tabular}{lccccccc}
            \hline
            \multicolumn{1}{l}{Model} & \multicolumn{7}{c}{Benchmark Signal}                                                                                                                                                                                                      \\
            \hline
            \multicolumn{1}{l}{}      & \multicolumn{1}{c}{\multirow{2}{*}{FCNC}} & \multicolumn{3}{c}{HG}        & \multicolumn{3}{c}{No HG}                                                                                                                                     \\
            \multicolumn{1}{l}{}      & \multicolumn{1}{c}{}                      & \multicolumn{1}{c}{$1.0$ TeV} & \multicolumn{1}{c}{$1.2$ TeV} & \multicolumn{1}{c}{$1.4$ TeV} & \multicolumn{1}{c}{$1.0$ TeV} & \multicolumn{1}{c}{$1.2$ TeV} & \multicolumn{1}{c}{$1.4$ TeV} \\
            \hline
            \multicolumn{8}{c}{Full features}                                                                                                                                                                                                                                     \\
            \hline
            Supervised DNN            & $2.9 ^{+1.4}_{-0.9}$                      & $0.09 ^{+0.04}_{-0.03}$       & $0.3 ^{+0.2}_{-0.1}$          & $0.17 ^{+0.07}_{-0.06}$       & $0.26 ^{+0.13}_{-0.08}$       & $1.9 ^{+1.3}_{-0.8}$          & $2.3 ^{+1.1}_{-0.7}$          \\
            $H_T$                     & $60 ^{+20}_{-20}$                         & $0.27 ^{+0.14}_{-0.09}$       & $0.3 ^{+0.2}_{-0.1}$          & $0.29 ^{+0.16}_{-0.09}$       & $0.8 ^{+0.5}_{-0.2}$          & $1.9 ^{+0.9}_{-0.7}$          & $3.2 ^{+1.7}_{-1.0}$          \\
            Deep SVDD                 & $6 ^{+3}_{-1}$                            & $0.4 ^{+0.1}_{-0.1}$          & $0.4 ^{+0.2}_{-0.1}$          & $0.5 ^{+0.2}_{-0.1}$          & $0.9 ^{+0.4}_{-0.3}$          & $2.6 ^{+1.2}_{-0.8}$          & $7 ^{+4}_{-2}$                \\
            AE                        & $20 ^{+4}_{-9}$                           & $0.25 ^{+0.13}_{-0.08}$       & $0.26 ^{+0.13}_{-0.08}$       & $0.28 ^{+0.13}_{-0.09}$       & $0.6 ^{+0.2}_{-0.2}$          & $1.4 ^{+0.7}_{-0.4}$          & $4 ^{+1}_{-1}$                \\
            HBOS                      & $60 ^{+20}_{-20}$                         & $0.3 ^{+0.2}_{-0.1}$          & $0.4 ^{+0.2}_{-0.1}$          & $0.4 ^{+0.2}_{-0.1}$          & $0.8 ^{+0.4}_{-0.2}$          & $2.2 ^{+1.0}_{-0.7}$          & $5 ^{+3}_{-1}$                \\
            iForest                   & $70 ^{+30}_{-20}$                         & $0.4 ^{+0.1}_{-0.2}$          & $0.4 ^{+0.2}_{-0.1}$          & $0.5 ^{+0.2}_{-0.2}$          & $0.9 ^{+0.4}_{-0.3}$          & $2.3 ^{+1.3}_{-0.7}$          & $6 ^{+4}_{-2}$                \\

            \hline
            \multicolumn{8}{c}{Sanitised features}

            \\    \hline
            Supervised DNN            & $2.8 ^{+1.3}_{-0.9}$                      & $0.22 ^{+0.18}_{-0.1}$        & $0.3 ^{+0.2}_{-0.1}$          & $0.4 ^{+0.2}_{-0.2}$          & $0.5 ^{+0.5}_{-0.2}$          & $1.8 ^{+1.4}_{-0.8}$          & $5 ^{+5}_{-2}$                \\

            $H_T$                     & $50 ^{+20}_{-10}$                         & $0.27 ^{+0.14}_{-0.09}$       & $0.3 ^{+0.16}_{-0.1}$         & $0.29 ^{+0.16}_{-0.09}$       & $0.8 ^{+0.5}_{-0.2}$          & $1.8 ^{+1.0}_{-0.5}$          & $3 ^{+2}_{-1}$                \\

            Deep SVDD                 & $6 ^{+3}_{-2}$                            & $0.19 ^{+0.08}_{-0.05}$       & $0.21 ^{+0.1}_{-0.05}$        & $0.24 ^{+0.11}_{-0.06}$       & $0.36 ^{+0.16}_{-0.09}$       & $1.1 ^{+0.5}_{-0.3}$          & $3.6 ^{+1.5}_{-0.9}$          \\
            AE                        & $60 ^{+30}_{-20}$                         & $0.9 ^{+0.5}_{-0.3}$          & $0.8 ^{+0.4}_{-0.3}$          & $0.6 ^{+0.4}_{-0.2}$          & $1.6 ^{+1.0}_{-0.5}$          & $4 ^{+2}_{-1}$                & $9 ^{+5}_{-3}$                \\
            HBOS                      & $60 ^{+20}_{-20}$                         & $0.5 ^{+0.3}_{-0.2}$          & $0.5 ^{+0.3}_{-0.2}$          & $0.5 ^{+0.3}_{-0.2}$          & $1.0 ^{+0.5}_{-0.4}$          & $2.4 ^{+1.5}_{-0.8}$          & $6 ^{+3}_{-2}$                \\
            iForest                   & $70 ^{+30}_{-20}$                         & $0.5 ^{+0.2}_{-0.2}$          & $0.5 ^{+0.2}_{-0.2}$          & $0.4 ^{+0.2}_{-0.1}$          & $1.1 ^{+0.5}_{-0.4}$          & $2.4 ^{+1.2}_{-0.8}$          & $5 ^{+3}_{-2}$                \\
        \end{tabular}
    \end{center}
\end{table}

\begin{figure}[h!]
    \centering
    \hspace*{-1cm}
    \includegraphics[width=0.75\textwidth]{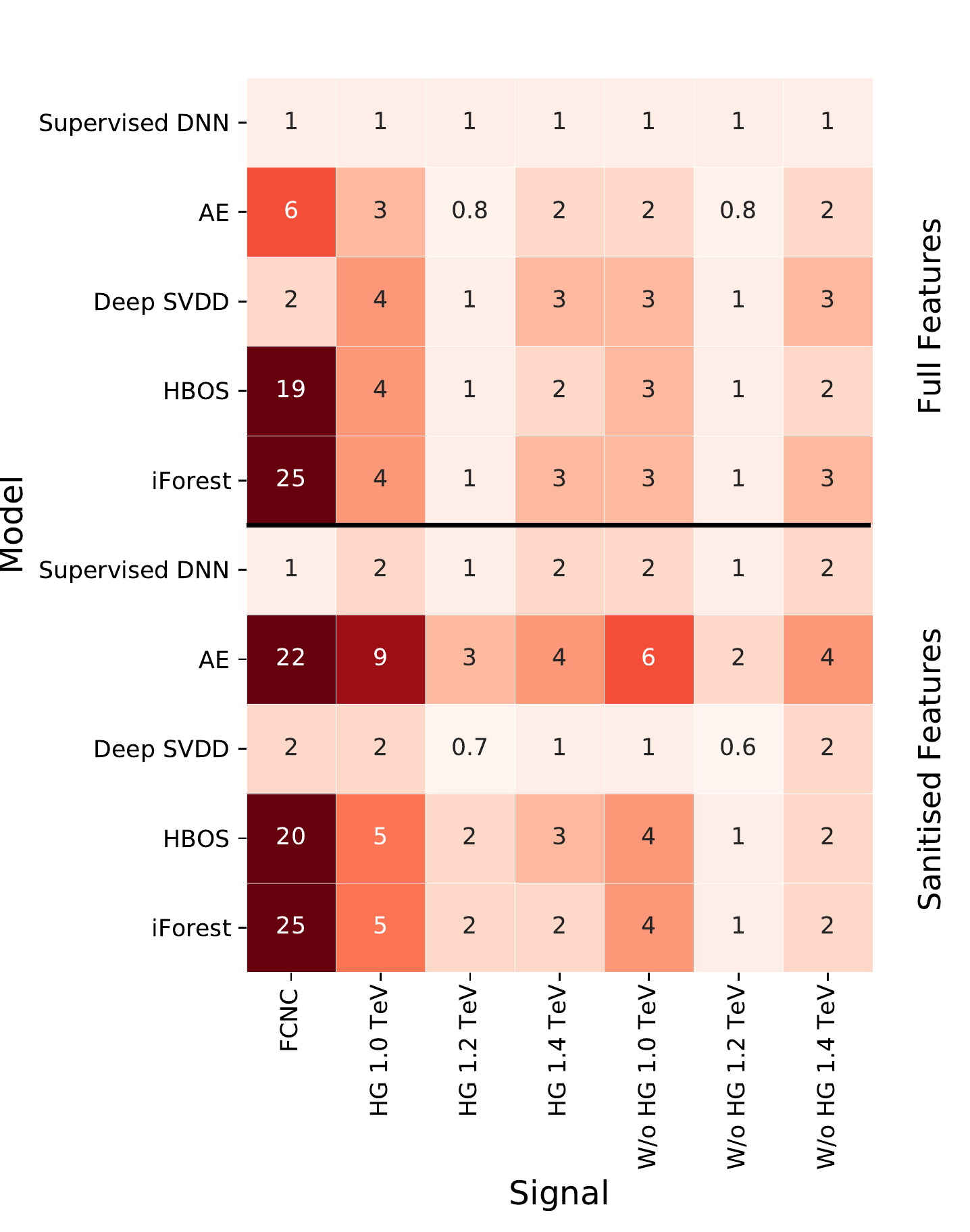}
    \caption{\label{fig:limits} 95\% CL upper limits on $\mu$ normalised to the limit obtained for the supervised DNN model.}
\end{figure}

We fit the distributions presented in~\cref{fig:fit_inputs}, to determine 
upper limits on the signal strength. In~\cref{tab:limits} we show the 
central values of the upper limit on $\mu$ and the associated statistical 
uncertainties. In \cref{fig:limits} are presented the same central values 
but normalised to the first line, \emph{i.e.} to the supervised DNN using 
the full feature set. We observe that the deep models, both AD and 
supervised, experienced performance impact by switching the feature set. In particular, we noticed 
how the Deep SVDD significantly improved when using the sanitised features 
for all cases, while the AE performance degraded. Furthermore, 
the Deep SVDDD has a sensitivity similar 
to supervised DNN across all signals when using the sanitised 
features. On the other 
hand, the shallow models retained the same discriminating power when 
changing the features.

Another relevant result that we observe is how, with sanitised features, the 
AE seems to focus more strongly on the out tails of the distributions in the 
same way as the shallow methods. On a different direction, the Deep SVDD 
produced similar discriminant power for all signals, including the FCNC, 
which is far more similar to the SM distribution than the signals with VLQ. 
This reinforces the idea that different AD algorithms are capturing outliers 
differently. Deep SVDD, for instance, might be interesting 
in searches for signals of new physics implying small deviations of the SM. 
A more detailed study of this behaviour, as well as of the propagation of 
systematic sources of uncertainties through these methods is left for a 
future study.

For comparison, \cref{tab:limits} also shows the limits for each signal type 
obtained by fitting the distribution of the scalar sum of transverse 
momentum ($p_T$) of all reconstructed particles in the event ($H_T$) as a 
simpler, but commonly used~\cite{ATLASZtag}, alternative to the use of ML 
methods. While the shallow methods are always worst than a simple $H_T$ 
fitting, the Deep SVDD seems to provide consistently better performance when 
using sanitised features across all signals, being better than a simple 
$H_T$ for the FCNC for both feature sets.

The results show that these unsupervised AD algorithms are reasonably sensitive to new signals, with a maximum degradation relative to the supervised DNN of around an order of magnitude on the $\mu$ exclusion limits, for the worst cases, and no significant impact for the best ones. Interestingly, in previous work where DNN trained on different models were used to discriminate between the background and other signals~\cite{TransfDL}, we observed similar trends when training deep neural networks on signals different from those used for the classification.

\FloatBarrier
\newpage

\section{Robustness of the anomaly detection methods}

\begin{figure}[h]
    \begin{center}
        \includegraphics[width=0.44\textwidth]{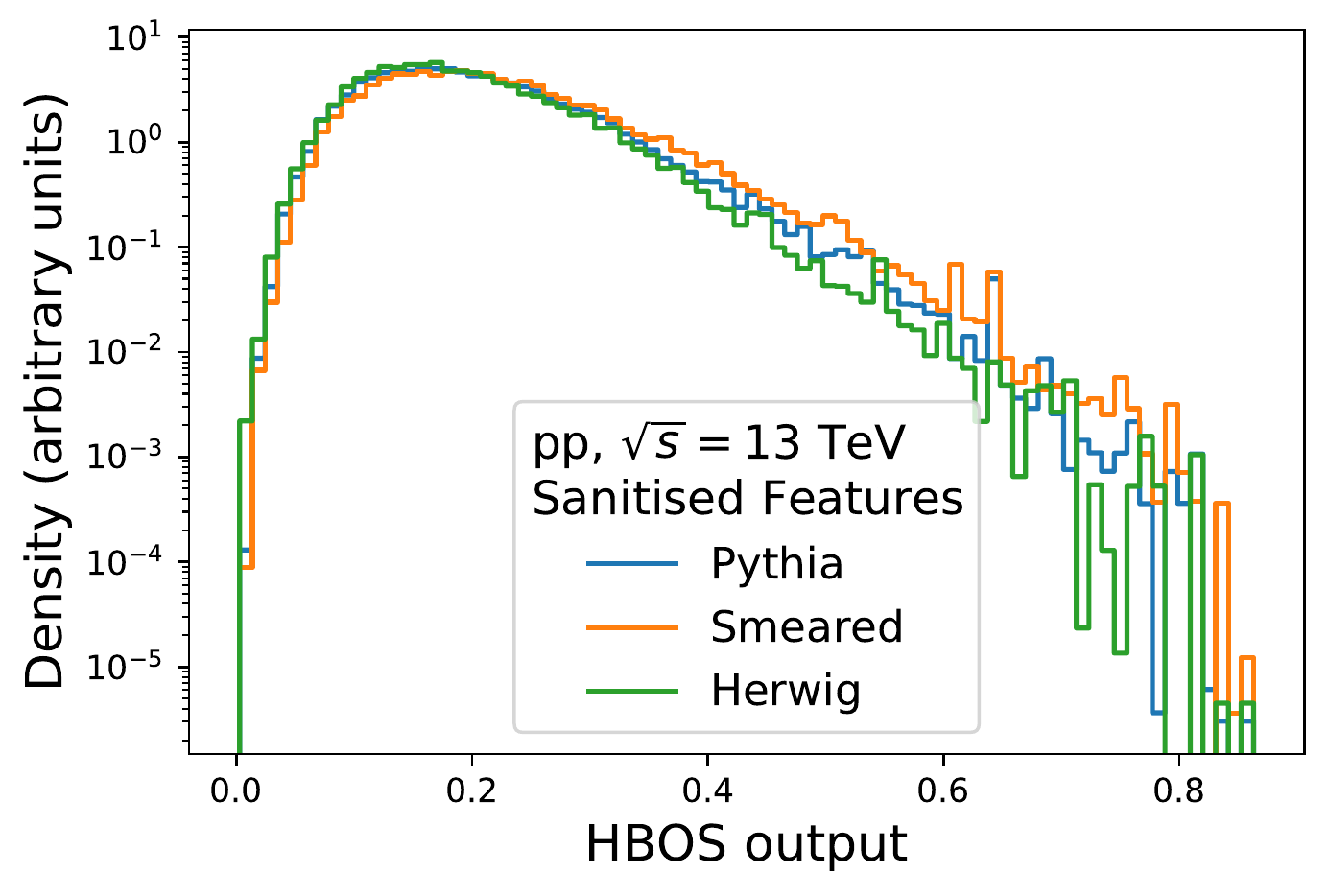}
        \includegraphics[width=0.44\textwidth]{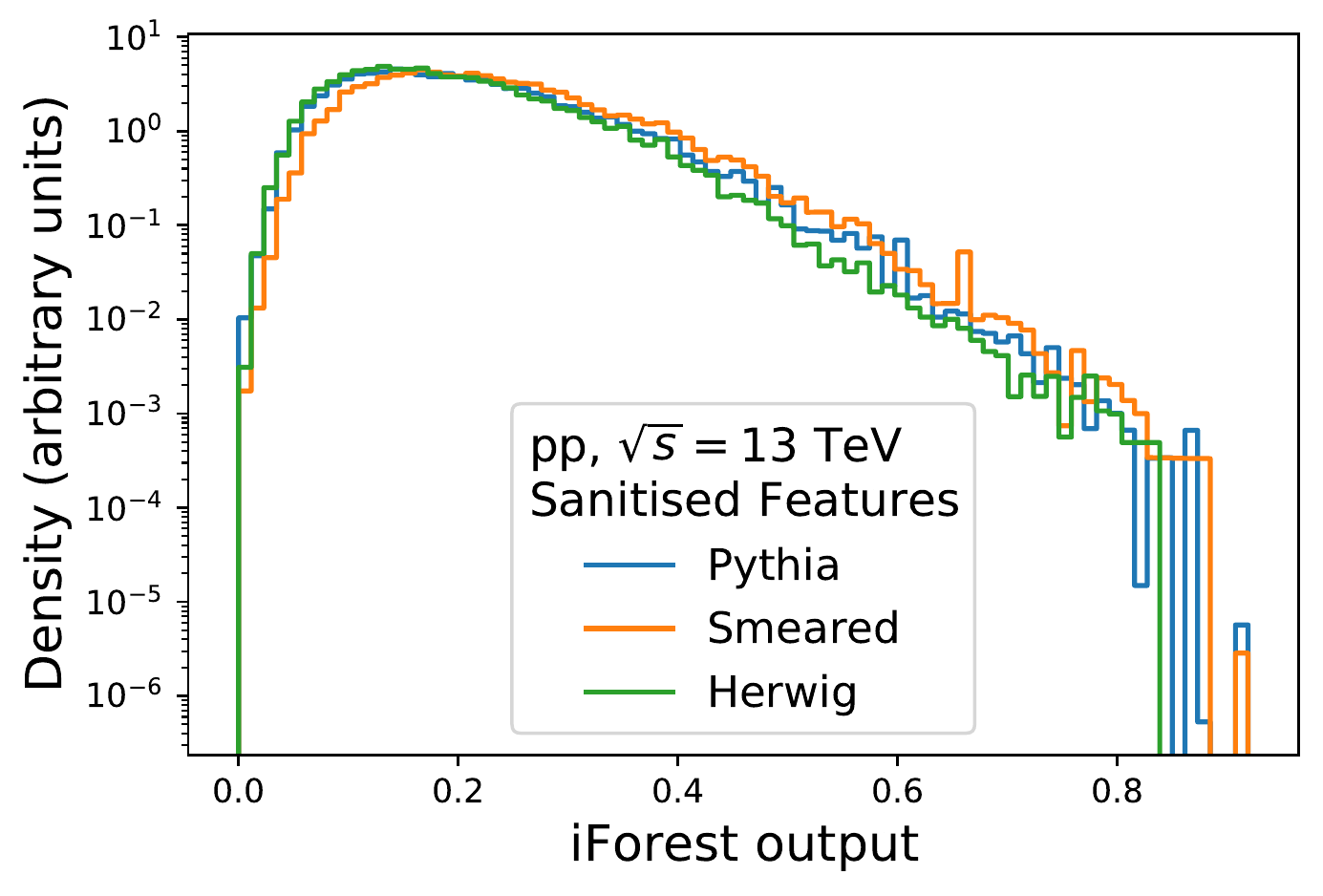}
        \includegraphics[width=0.44\textwidth]{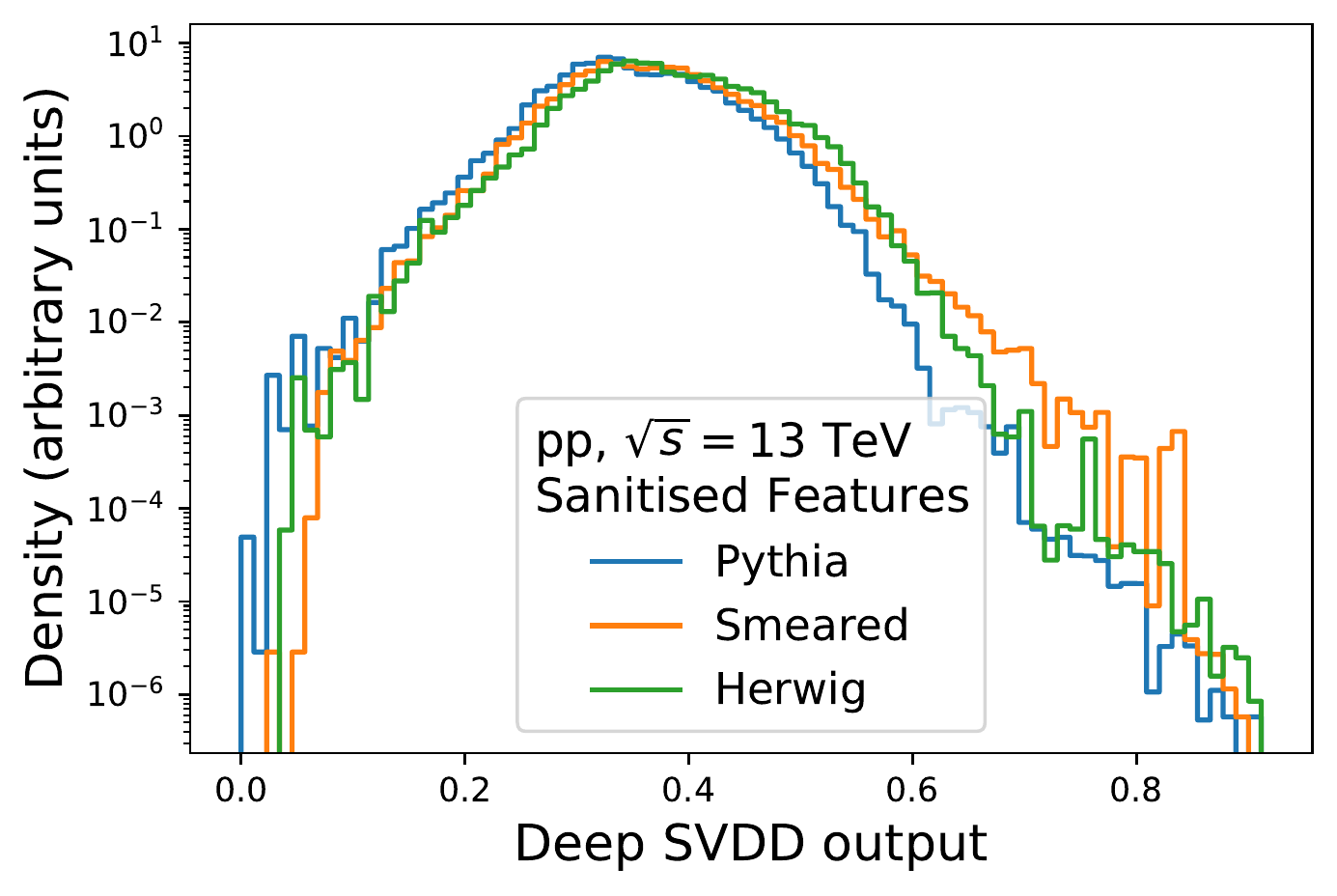}
        \includegraphics[width=0.44\textwidth]{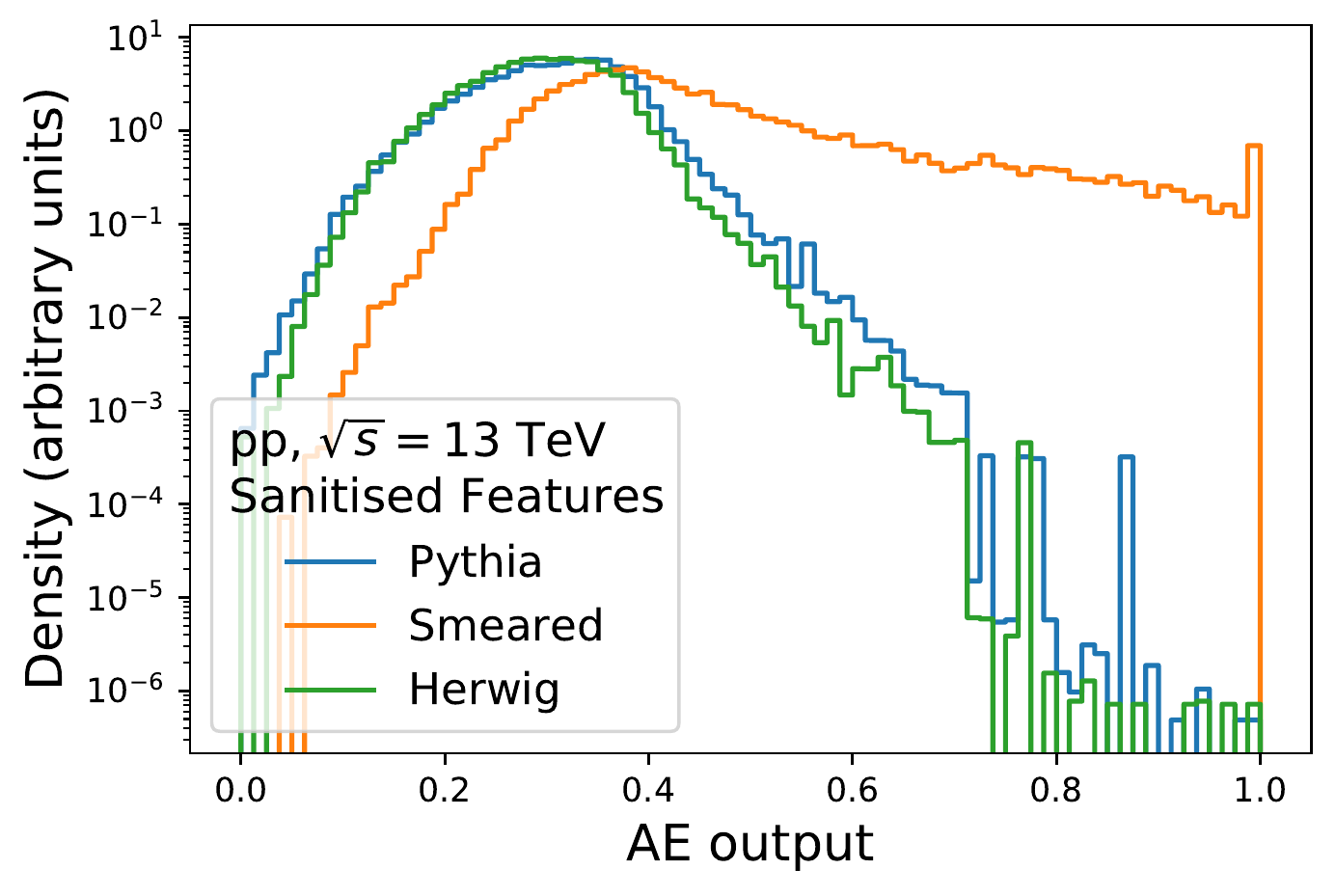}

        \caption{\label{fig:pythia_herwig_smeared_outputs}
            Anomaly score for the different AD methods (HBOS, iForest, Autoencoder, DeepSVDD) for the sanitised features on the test data of the original sample (Pythia), $p_T$ smeared sample (Smeared), and Herwig sample (Herwig).  The distributions are normalised to the unit area.
        }
    \end{center}
\end{figure}

In order to study the robustness of the presented models against background mismodelling we performed two simple experiments. In the first experiment, we smeared the $p_T$ of all objects with a Gaussian noise with standard deviation of $0.1$. For the second experiment, we switch the hadroniser from Phytia to Herwig, whilst maintaining everything else the same. The outputs of the AD models trained on the original Pythia sample with the sanitised features for both cases are presented in~\cref{fig:pythia_herwig_smeared_outputs}.

For the $p_T$ smeared test, we observe that the mass of the output distribution of each AD model is shifted to the right, meaning that the new sample is deemed more anomalous than the original Pythia sample. More interestingly, we observe a considerable change in shape of the output distribution for the AE, suggesting that this method is specially sensitive to mismodelling of the $p_T$. On the other hand, the Deep SVDD seems  more robust to this smear, although it pushes some background to the region where one would expect signal. Finally, we notice that the shallow methods, being simpler, are clearly more robust against $p_T$ smearing. In addition, we derived the expected upper limits with the smeared $p_T$ using the AD models trained on the original sample. We observed that the shallow methods produced values of $\mu$ compatible with those presented in~\cref{tab:limits} within the statistical uncertainty, while the results  for deeper models got worse. The central values for $\mu$ for the Deep SVDD increased on average 2 to 3 times, while still maintaining exclusion power. For the AE, however, the limits worsen by two orders of magnitude, as one would expect from~\cref{fig:pythia_herwig_smeared_outputs}.

For the Herwig sample we notice, once again, that the output distributions for the shallow methods are considerably less modified, while they differ from the expected Pythia output distributions for the deep methods. For the deep methods we observe different effects. While for the Deep SVDD the distribution moves to the right, for the AE it seems to move to the left. As before, we produced expected upper limits using the Herwig samples and compared them to the ones obtained using the original Pythia sample in~\cref{tab:limits}. Just like with the $p_T$ smeared case, the shallow methods proved to be the more robust with $\mu$ values compatible with the ones derived with the Pythia sample. For the deep methods, the Deep SVDD produced limits around twice as large as for the Pythia sample, but still with smaller degradation than those obtained with the $p_T$ smeared case. For the AE we observed a degradation as severe as with the $p_T$ smeared case, with the limits worsening by two orders of magnitude.

These two tests suggest that the methods presented in this work can be sensitive to mismodelling, and point to the need for a thorough study of the impact of systematic uncertainties and how to mitigate the effect of such uncertainties on the sensitivity to new phenomena beyond the Standard Model. Such comprehensive study is outside the scope of the presented work.

\section{Conclusions}

In this work, we studied four distinct unsupervised AD algorithms, two 
shallow and two deep, which were trained on simulated SM events. The 
resulting trained models provided us with an anomaly score that was then 
used to perform upper bounds on seven benchmark signals covering three 
classes of new physics: FCNC interaction, SM gluon VLQ production, and heavy 
gluon VLQ production. Even though all algorithms eventually targeted events 
at the tails of the original SM distributions, they capture different events 
and are therefore learning different notions of \emph{outlyingness}. This 
was clearly observed on how the Deep SVDD and the AE performed between VLQ 
and FCNC signals. Upper limits on the signal strength were obtained by 
fitting the output distributions of each AD model using the $CL_s$ method. 
We showed that the deep models, namely the Deep SVDD, can outperform the 
shallow ones, and each deep model performed differently depending on the 
broader class of signals being tested. This result suggests that different 
AD algorithms are suitable to isolate different types of BSM physics and are 
complementary to each other in unsupervised generic searches for new 
physics.

\section{Acknowledgments}

We thank Guilherme Milhano, Maria Ramos and Guilherme Guedes for the careful reading of the manuscript and for the useful discussions. We also thank Ana Peixoto and Tiago Vale for providing the MadGraph cards used for the simulation of the beyond the Standard Model samples. We acknowledge the support from FCT Portugal,
Lisboa2020, Compete2020, Portugal2020 and FEDER under project
PTDC/FIS-PAR/29147/2017. The computational part of this work was supported by
INCD (funded by FCT and FEDER under the project 01/SAICT/2016 nr.
022153) and by the Minho Advanced Computing Center (MACC). The Titan Xp
GPU card used for the training of the Deep Neural Networks developed for
this project was kindly donated by the NVIDIA Corporation.

\bibliography{paper}{}
\bibliographystyle{unsrt}

\end{document}